{}

\documentclass[11pt,a4paper]{article}

\newif\ifpublic
\publictrue

\usepackage[bookmarks=true,hyperfigures=true,colorlinks=true,linkcolor=black,citecolor=black,urlcolor=black,bookmarksnumbered]{hyperref}

\usepackage[nosort]{cite}

\usepackage[parsep]{collref}
\usepackage[english]{babel}
\usepackage[T1]{fontenc}
\usepackage[latin1]{inputenc}
\usepackage{amsfonts,amsbsy,dsfont,euscript,mathrsfs,fixmath}
\usepackage{amsmath,amssymb}
\usepackage{enumerate}
\usepackage{slashed}
\ifpublic\else\usepackage{showkeys}\fi

\usepackage[a4paper,text={160mm,247mm},centering]{geometry}
\allowdisplaybreaks[3]
\numberwithin{equation}{section}
\def\[{\begin{equation}\begin{aligned}}
\def\]{\end{aligned}\end{equation}}
\newcommand{\nn}{\nonumber}

\usepackage[font=small,labelfont=bf,width=0.85\textwidth]{caption}

\expandafter\def\expandafter\bfseries\expandafter{\bfseries\ifmmode\else\boldmath\fi}
\expandafter\def\expandafter\mdseries\expandafter{\mdseries\ifmmode\else\unboldmath\fi}
\expandafter\def\expandafter\normalfont\expandafter{\normalfont\ifmmode\else\unboldmath\fi}

\RequirePackage{verbatim}

\makeatletter
\newwrite\bibinl@out
\newenvironment{bibtex}[1][\jobname]{%
\immediate\openout\bibinl@out #1.bib%
\immediate\write\bibinl@out{\@percentchar generated from `\jobname' starting line \the\inputlineno^^J}%
\def\verbatim@processline{\immediate\write\bibinl@out{\the\verbatim@line}}%
\@bsphack\let\do\@makeother\dospecials\catcode`\^^M\active\verbatim@start%
}
{\immediate\closeout\bibinl@out\@esphack}
\makeatother

\let\barefrac=\frac
\renewcommand{\frac}[2]{\mathinner{\barefrac{#1}{#2}}}

\let\baresqrt=\sqrt
\makeatletter
\renewcommand{\sqrt}{\@ifnextchar[\@sqrt@space@a\@sqrt@space@b}
\def\@sqrt@space@a[#1]#2{\mathinner{\mathchoice{\mkern-3mu}{\mkern-3mu}{}{}\baresqrt[#1]{#2}}}
\def\@sqrt@space@b#1{\mathinner{\mathchoice{\mkern-3mu}{\mkern-3mu}{}{}\baresqrt{#1}}}
\makeatother

\makeatletter
\let\per@dot@old=\.
\def\.{\ifmmode\def\per@dot@sel{\mkern3mu}\else\def\per@dot@sel{\per@dot@old}\fi\per@dot@sel}
\makeatother

\let\barefootnote=\footnote
\renewcommand{\footnote}[1]{\barefootnote{#1\vspace{3pt}}}

\newcommand{\vfrac}[2]{\ifmmode\mathinner{\textstyle^{#1}\!/\!_{#2}}\else$^{#1}\!/\!_{#2}$\fi}

\DeclareMathOperator{\diag}{diag}
\DeclareMathOperator{\tr}{tr}

\DeclareMathOperator{\STr}{STr}

\newcommand{\Real}{\mathds{R}}

\newcommand{\Integer}{\mathds{Z}}

\let\Re\relax\DeclareMathOperator{\Re}{Re}
\let\Im\relax\DeclareMathOperator{\Im}{Im}

\newcommand{\ind}[1]{{\scriptscriptstyle{#1}}}

\newcommand{\alg}[1]{\mathfrak{#1}}
\newcommand{\grp}[1]{\mathrm{#1}}

\DeclareMathOperator{\Ad}{Ad}

\newcommand{\com}[2]{[#1,#2]}
\newcommand{\anticom}[2]{\{#1,#2\}}

\def\<{\big\langle}
\def\>{\big\rangle}

\newcommand{\geom}[1]{\mathrm{#1}}
\newcommand{\Man}{\geom{M}}
\newcommand{\AdS}{\geom{AdS}}

\newcommand{\Sp}{\geom{S}}

\newcommand{\Hy}{\geom{H}}
\newcommand{\To}{\geom{T}}

\newcommand{\extder}{\mathrm{d}}

\def \a {\alpha}
\def \adss {$\AdS_3 \times \Sp^3$}

\providecommand{\href}[2]{#2}

\makeatletter
\def\mr@ignsp#1 {\ifx\:#1\@empty\else #1\expandafter\mr@ignsp\fi}
\newcommand{\multiref}[1]{\begingroup%
\xdef\mr@no@sparg{\expandafter\mr@ignsp#1 \: }%
\def\mr@comma{}\def\mr@dash{-}%
\@for\mr@refs:=\mr@no@sparg\do{%
\ifx\mr@refs\mr@dash\def\mr@comma{}--\else%
\mr@comma\def\mr@comma{,}\ref{\mr@refs}\fi}%
\endgroup}
\renewcommand{\eqref}[1]{(\multiref{#1})}
\makeatother

\makeatletter
\newcommand{\namedref}[2]{\hyperref[#2]{#1~\ref*{#2}}}
\newcommand{\secref}{\@ifstar{\namedref{Section}}{\namedref{sec.}}}
\newcommand{\appref}{\@ifstar{\namedref{Appendix}}{\namedref{app.}}}
\newcommand{\tabref}{\@ifstar{\namedref{Table}}{\namedref{tab.}}}
\newcommand{\figref}{\@ifstar{\namedref{Figure}}{\namedref{fig.}}}
\makeatother

\let\oldbib=\thebibliography
\def\thebibliography{\phantomsection\addcontentsline{toc}{section}{\refname}\oldbib}

\providecommand{\hypersetup}[1]{}
\providecommand{\texorpdfstring}[2]{#1}
\hypersetup{plainpages=false}
\hypersetup{pdfpagemode=UseNone}
\hypersetup{bookmarksnumbered=true}
\hypersetup{pdfstartview=FitH}
\hypersetup{colorlinks=false}
\hypersetup{citebordercolor={1 1 1}}
\hypersetup{urlbordercolor={1 1 1}}
\hypersetup{linkbordercolor={1 1 1}}

\makeatletter
\let\@keywords\@empty
\let\@subject\@empty
\providecommand{\keywords}[1]{\gdef\@keywords{#1}}
\providecommand{\subject}[1]{\gdef\@subject{#1}}
\def\thetitle{\@title}
\def\theauthor{\@author}
\def\thesubject{\@subject}
\def\thedate{\@date}
\def\thekeywords{\@keywords}
\makeatother
\AtBeginDocument{
\hypersetup{pdftitle={\thetitle}}
\hypersetup{pdfauthor={\theauthor}}
\hypersetup{pdfsubject={\thesubject}}
\hypersetup{pdfkeywords={\thekeywords}}}

\newif\ifshownote
\shownotetrue

\ifshownote

\ifpdf\else\RequirePackage[active]{srcltx}\fi
\RequirePackage{xcolor}
\RequirePackage{ulem}

\newcommand{\remark}[2][]{{\normalfont\sffamily\hspace{1ex}
\def\emph{\textsl}\def\textbullet{$\bullet$}
\def\tmparga{#1}
\def\tmpargb{BH}\ifx\tmparga\tmpargb\color[rgb]{0.5,0,0}\fi
\def\tmpargb{FS}\ifx\tmparga\tmpargb\color[rgb]{0,0.5,0}\fi
\def\tmpargb{}\ifx\tmparga\tmpargb\color{red}\fi
\def\tmpargb{}\ifx\tmparga\tmpargb\else \textbf{#1:}\fi
#2\hspace{1ex}}}

\newcommand{\rf}[1]{(\ref{#1})}

\else
\newcommand{\remark}[2][]{\ignorespaces}

\fi

\def \N {{\cal N}}
\def \ka {\kappa}
\def \ha {\tfrac{1}{2}}\def \ci {\cite} \def \la{\label}
\def \ed{ \bibliographystyle{nb} \bibliography{Refs}
\end{document}}

 \def \k {\kappa}  \def \k {\kappa}
\def \adss {$\rm AdS_3 \times S^3$ } \def \adst {$\rm AdS_2 \times S^2$ } \def \N {{\cal N}}
\def \iffa {\iffalse} \def \k {\kappa}
\def \no {\nonumber} \def \foot {\footnote} 
\def \td {\tilde}
\def \CC {{ \check B}}
\def \P {\Phi}
\def \be {\begin{equation}} \def \ee {\end{equation}}
\def \x {\xi}\def \s {\theta}
\def \CCC {{\td{\check B}}}
\def \bz {\textbf{z}} \def \dd {\extder} \def \g {\gamma} 
\def \adsk {$\AdS_{3\k} \times \Sp^3_\k$ }
\def \nn {\no \\ }

  \def \x {\xi} \def \bs {\sigma}

\def \m {\mu}
\def \RR {{\cal R}}
\def \ep {\epsilon}
\def \vk {\varkappa}
\def \rz {\mathrm{z}}
\def \bF {{\bf F}}
\def \z {\zeta}

\title{Integrable supersymmetric deformations of \texorpdfstring{$\AdS_3 \times \Sp^3 \times \To^4$}{AdS3 x S3 x T4}}

\begin{document}

\begin{flushright}\small{Imperial-TP-AT-2022-{02}}
\end{flushright}
\vspace{1.5cm}

\begin{center}

{\Large\bf Integrable supersymmetric deformations of $\AdS_3 \times \Sp^3 \times \To^4$ \\
\vspace{0.3cm}
}

\vspace{1.0cm}

{Ben Hoare$^{a,}$\footnote{ben.hoare@durham.ac.uk}, \ Fiona K. Seibold$^{b,}$\footnote{f.seibold21@imperial.ac.uk} and
Arkady A. Tseytlin$^{b,}$\footnote{Also on leave from Institute for Theoretical and Mathematical Physics (ITMP) and
Lebedev Institute.
\\ \hspace*{0.5cm} tseytlin@imperial.ac.uk}
}

\vspace{0.5cm}

{
\vspace{0.15cm}
$^{a}$ \it Department of Mathematical Sciences,
Durham University,
Durham DH1 3LE, UK\\
\vspace{0.05cm}
\vspace{0.15cm}
$^{b}$\it Blackett Laboratory, Imperial College, London SW7 2AZ, UK
}
\end{center}

\vspace{0.5cm}

\begin{abstract}
We construct a family of type IIB string backgrounds that are deformations of $\rm AdS_3 \times S^3 \times T^4$
with a ``squashed'' $\rm AdS_3 \times S^3$ metric supported by a combination of NSNS and RR fluxes.
They have global $\grp{SU}(1,1) \times \grp{SU}(2)$ symmetry, regular curvature, constant dilaton and preserve 8
supercharges. Upon compactification to 4 dimensions they reduce to $\N=2$ supersymmetric
$\rm AdS_2 \times S^2$ solutions with electric and magnetic Maxwell fluxes.
These type IIB supergravity solutions can be found from the undeformed $\rm AdS_3 \times S^3 \times T^4$ background by a combination
of T-dualities and S-duality.
In contrast to T-duality, S-duality transformations of a type IIB supergravity background do not generally preserve the classical integrability of the corresponding Green-Schwarz superstring sigma model.
Nevertheless, we show that integrability is preserved in the present case.
Indeed, we find that these backgrounds can be obtained, up to T-dualities,
from an integrable inhomogeneous Yang-Baxter deformation (with unimodular Drinfel'd-Jimbo R-matrix)
of the original
$\AdS_3 \times \Sp^3$ supercoset model.
\end{abstract}

\newpage

\tableofcontents

\setcounter{footnote}{0}
\setcounter{section}{0}
\section{Introduction}

Exact solutions of classical string theory that have a direct target space interpretation (i.e.~are described by conformal sigma models)
are rare and hard to find.
For most relevant leading-order solutions, such as non-supersymmetric 4d black holes, their exact form is not known.
Additional global symmetries (in particular, supersymmetry)
are important to have some control over deformations induced by $\a'$-corrections.
An even smaller subclass of string sigma models are integrable (and thus have, in principle, a solvable string spectrum).

Given an integrable background that has some isometries one can generate new integrable solutions with more parameters
by T-dualities (for some examples see, e.g., \cite{Russo:1995tj,Tseytlin:1995fh,Maldacena:1999mh,Lunin:2005jy,Frolov:2005ty,Frolov:2005dj,Alday:2005ww,Orlando:2019rjg}).
On the other hand, S-duality maps one type IIB supergravity solution into another,
but is not a symmetry of the classical string theory (it does not act on the string worldsheet) and thus, in contrast to T-duality,
does not, in general, ``commute'' with $\a'$-corrections or integrability.

Supersymmetric integrable string backgrounds like $\AdS_n \times \Sp^n \times \To^{10-2n}$
that appear as near-horizon limits of brane configurations are of particular interest.
It is important to study closely related solutions with non-trivial parameters
that are also supersymmetric and integrable. Having extra parameters may help
clarify the structure of the underlying integrable S-matrix and ``resolve'' special limits.

Below we will present a new 8-parameter class of
deformed $\AdS_3 \times \Sp^3 \times \To^{4}$
type IIB backgrounds supported by a combination of homogeneous NSNS and RR fluxes.
They have global $\grp{SU}(1,1) \times \grp{SU}(2)$ symmetry, regular curvature, constant dilaton and
preserve $\tfrac{1}{4}$ of maximal 10d supersymmetry. As type IIB supergravity solutions their existence may not be surprising -- they can be obtained
from undeformed $\AdS_3 \times \Sp^3 \times \To^{4}$ (supported by RR 3-form flux)
by a combination of T-dualities and S-dualities.
What is non-trivial is that the corresponding
Green-Schwarz (GS) superstring sigma model
will be also integrable.

As the relation between the undeformed and deformed backgrounds will involve S-duality,
integrability
will not simply follow from the known integrability of the original
$\AdS_3 \times \Sp^3 \times \To^{4}$ model (see \ci{Babichenko:2009dk,Sfondrini:2014via,OhlssonSax:2018hgc} and references there).
The proof of integrability will be based on the key observation that a particular subclass
of backgrounds (from which the others can be obtained by just T-dualities)
correspond to a Yang-Baxter (YB) deformed supercoset model \ci{Delduc:2018xug} with a particular Drinfel'd-Jimbo R-matrix
\ci{Hoare:2018ngg,Seibold:2019dvf}.
Being an inhomogeneous YB deformation it will not simply be equivalent to a T-duality transformation of the
original background.
Also, in contrast to some familiar examples of YB or $\eta$-deformations (see, e.g.,
\ci{Hoare:2014pna,Hoare:2021dix} for a review)
that have few manifest symmetries, singularities and solve the generalized supergravity equations
\ci{Arutyunov:2015qva,Arutyunov:2015mqj,Wulff:2016tju}, here the resulting background
will share the key features of the undeformed \adss --
manifest non-abelian isometry, supersymmetry, regular curvature, constant dilaton and, most importantly,
will solve the standard type IIB
supergravity equations, i.e.~will represent a consistent string model.

\medskip

Let us recall that the standard
$\AdS_3 \times \Sp^3 \times \To^4$ background can be supported by a mix of NSNS and RR 3-form fluxes.
In the pure NSNS case the worldsheet theory is a supersymmetric extension of the $\grp{SL}(2,\Real)\times\grp{SU}(2)$ WZW theory
(and thus admits a local NSR description and is solvable by 2d CFT methods).
The model with non-zero RR flux has a local GS description and its integrability
follows from its construction as a
sigma model on the semi-symmetric
supercoset $\grp{G}/\grp{H}$ with $\grp{G} = \grp{PSU}(1,1|2) \times \grp{PSU}(1,1|2)$ and $\grp{H}=\grp{SU}(1,1) \times \grp{SU}(2)$
\ci{Rahmfeld:1998zn,Park:1998un,Berkovits:1999im,Metsaev:2000mv,Babichenko:2009dk,Wulff:2014kja}.

We will be interested in ``warped'' or ``squashed''
deformations of $\AdS_3 \times \Sp^3 \times \To^4$ (depending on a
continuous deformation parameter $\kappa$) which preserve only
half of the global symmetries,\foot{The isometry group of undeformed
$\AdS_3$ is $\grp{SU}(1,1) \times \grp{SU}(1,1)$, while the one of $\Sp^3$ is $\grp{SU}(2) \times \grp{SU}(2)$.}
i.e.~$\grp{SU}(1,1) \cong \grp{SL}(2,\Real)$ and $\grp{SU}(2)$. One way to obtain such deformed $\AdS_{3 \k}$ and $\Sp^3_{\k}$ geometries is to apply TsT transformations involving a particular abelian isometry
as well as one extra torus direction.\foot{Let us mention that
marginal NSNS ``${\rm J}\bar {\rm J}$'' deformations
\cite{Hassan:1992gi,Giveon:1993ph} of
the $\grp{SL}(2,\Real) \times \grp{SU}(2)$ WZW model generated by T-dualities were discussed, e.g., in \cite{Cvetic:1998vb,Manvelyan:2002xx}.
We also note that the ``squashed'' $\Sp^3$ sigma model was first shown to be integrable in \cite{Cherednik:1981df}.
This model with WZ term added is not conformal (has 2-parameter RG flow
in \cite{Demulder:2017zhz,Schubring:2020uzq,Levine:2021fof}) and is also integrable
\ci{Kawaguchi:2010jg,Kawaguchi:2011mz}.}
Applying this TsT transformation at the level of the GS model
generates integrable embeddings into type IIB string theory~\cite{Orlando:2010ay,Orlando:2012hu}.\foot{T-duality along Hopf fibres
was originally discussed in a similar context in \ci{Duff:1998cr}.}
For ${\rm AdS}_{3 \k} \times \Sp^3 \times \To^4$
and $\AdS_3 \times {\rm S}^3_{\k} \times \To^4$ the corresponding supergravity backgrounds preserve 8 supersymmetries,
while combining the two TsT transformations
leads to ${\rm AdS}_{3\k} \times {\rm S}^3_{\k'} \times \To^4$ that should break all supersymmetries \cite{Orlando:2010ay}.

Here we will find a different integrable embedding of the ${\rm AdS}_{3\k} \times {\rm S}^3_\k \times \To^4$
metric (with equal deformation parameters $\k$) into type IIB supergravity
that will preserve 8 supersymmetries and will be supported by a 7-parameter family
of homogeneous
NSNS and RR 3-form and 5-form fluxes.\foot{For examples when the same deformed metric can be supported by different
combinations of fluxes see, e.g., \cite{Lunin:2014tsa}.}
As mentioned above, it will not be related to the undeformed
$\AdS_3 \times \Sp^3 \times \To^4$ just by T-dualities and thus its integrability will be non-trivial.

From a broader perspective, our results are of interest in the context of the following questions:

(i) Given an integrable bosonic sigma model,
when is its embedding into superstring theory, with non-zero RR fluxes required for conformality, also integrable?\foot{If a GS sigma model is
conformal (or at least scale
invariant), hence it has $\varkappa$-symmetry \ci{Arutyunov:2015mqj,Wulff:2016tju},
one might think that $\vk$-symmetry implies integrability in the fermionic sector as well.
This need not, however, be true as the $\vk$-symmetry is a gauge redundancy -- fixing it will not, in general, leave
a symmetry relating bosons and fermions.}
It appears that a sufficient condition
for a positive answer is the preservation of a sufficient amount of target space supersymmetry:\foot{This is
not a necessary condition:
there are examples of integrable bosonic models that can be embedded into integrable superstring sigma models without any target space supersymmetry. These include the $\eta$- \cite{Delduc:2013qra} and $\lambda$- \cite{Hollowood:2014qma} deformations of the $\AdS_5 \times \Sp^5$ superstring, as well as the 3-parameter $\gamma$-deformed background \cite{Frolov:2005dj,Frolov:2005iq,Alday:2005ww}.
Even in these examples, the global supersymmetry of the undeformed theory is not completely lost however, since it effectively becomes ``hidden'' in the deformed one.} 
it may promote the classical integrability of the bosonic sector (the existence of Lax pair) to the
full GS model.
While a general proof of this is not known,
our family of backgrounds provides
a new explicit example of this connection (complementing the familiar ones discussed in
\ci{Bena:2003wd,Adam:2007ws,Wulff:2017vhv,Zarembo:2017muf}).

(ii) When do S-duality transformations of a background accidentally preserve the
integrability of the corresponding type IIB GS sigma model?
Here we find a new non-trivial example of this in addition to the familiar S-dual undeformed backgrounds
$\AdS_3 \times \Sp^3 \times \To^4$ with NSNS vs RR 3-form fluxes and
also to the ``Jordanian'' YB deformation ones discussed in \ci{Matsumoto:2014ubv,vanTongeren:2019dlq}.

\medskip

The structure of the rest of this paper is as follows.
We start in section \ref{sec:2} with the simplest example of the deformed ${\rm AdS}_{3\k} \times {\rm S}^3_\k $
background supported by a combination of NSNS and RR 3-form fluxes obtained by a TsT transformation in the two Hopf fibres of the undeformed
${\rm AdS}_{3} \times {\rm S}^3 $. We shall then describe the type IIB supergravity solutions where
the deformed ${\rm AdS}_{3\k} \times {\rm S}^3_\k \times \To^4$ metric is supported by a 7-parameter family
of NSNS $H_3$ and RR $F_3$ and $F_5$ fluxes (and constant scalars). We shall explain
how these backgrounds
can be obtained from the standard undeformed ${\rm AdS}_{3} \times {\rm S}^3 \times \To^4$ solution with $F_3$ flux
by U-duality, i.e.~by a combination of TsT and S-duality transformations. We shall highlight some special ``seed'' solutions, in particular, the one that has only non-vanishing $F_5$ flux.

Next, in section \ref{sec:susy} we will show that our ${\rm AdS}_{3\k} \times {\rm S}^3_\k \times \To^4$
solutions admit 8 Killing spinors, i.e.~preserves $\tfrac14$ of maximal 10d supersymmetry.
Section \ref{sec:int} will be devoted to demonstrating the integrability of the GS sigma model corresponding to a non-trivial
representative of the deformed backgrounds (from which all others can be obtained just by T-dualities).
We shall consider a particular YB deformation of
$\frac{\grp{PSU}(1,1|2) \times \grp{PSU}(1,1|2)}{\grp{SU}(1,1) \times \grp{SU}(2)} $ supercoset model
based on the Drinfel'd-Jimbo R-matrix built from a Cartan-Weyl basis with all fermionic simple roots.
In this case the solution of the modified classical Yang-Baxter equation is unimodular \ci{Hoare:2018ngg,Seibold:2019dvf}
and thus should lead \ci{Borsato:2016ose} to backgrounds that solve the
standard supergravity equations (rather than the generalized ones \ci{Arutyunov:2015qva,Arutyunov:2015mqj}).

In section \ref{sec:cont} we first consider the analytically-continued family of solutions with $\k= i \td\k$
in which the AdS$_3$ part is written as a Hopf fibration over the Minkowski-signature AdS$_2$ space.
The resulting type IIB background is shown to interpolate between
${\rm AdS}_{3} \times {\rm S}^3 \times \To^4$ (for $\td \k=0$) and ${\rm AdS}_{2} \times {\rm S}^2 \times \To^6$ (for $\td \k=1$), with the latter being supported by the $F_5$ flux only.
Dimensionally reducing
on the two Hopf fibres for any value of $\td \k$
we get a family of 4d supergravity solutions with ${\rm AdS}_{2} \times {\rm S}^2$ metric supported by a combination of
equal-charge electric and magnetic Maxwell fluxes that are familiar near-horizon limits of a family of
$\N=2, d=4$ BPS black holes (with constant scalars).
We shall also discuss special limits of our family of solutions including
a Schr\"odinger background corresponding to a particular
Jordanian limit of the YB deformation with Drinfel'd-Jimbo R-matrix.
We also construct the pp-wave limit, which describes the quadratic approximation
to the BMN-expanded superstring action and we comment on the corresponding dispersion relation and tree-level bosonic S-matrix for the corresponding string fluctuations.

Some concluding remarks will be made in section \ref{con}. Appendix \ref{susy} contains
details of the proof of supersymmetry in section \ref{sec:susy} and also
a construction of Killing spinors in the pp-wave limit. The background of an inhomogeneous YB deformation constructed using a non-unimodular Drinfel'd-Jimbo R matrix built from a distinguished Cartan-Weyl basis is presented in Appendix \ref{app:dist}. In contrast to the background discussed in section \ref{sec:int} it only solves a set of generalised supergravity equations of motion.
We also include some related integrable deformations with non-constant dilaton constructed using other TsT transformations in Appendix \ref{else}.

\section{Deformed \texorpdfstring{$\AdS_3 \times \Sp^3 \times \To^4$}{AdS3 x S3 x T4} backgrounds}
\label{sec:2}

In this section we will construct a class of $\Man^6 \times \To^4$ type IIB supergravity solutions where
the metric of $\Man^6$= \adsk will be that of ``squashed'' or
deformed \adss with the same deformation parameter $\k$ in the two factors.
These backgrounds will have regular curvature, constant dilaton,
homogeneous NSNS and RR 3-form and 5-form fluxes and global $\grp{SU}(1,1) \times \grp{SU}(2)$ symmetry. They will preserve 8
supercharges (see section \ref{sec:susy}). Moreover, as in the undeformed \adss case, the
corresponding GS superstring sigma model will be integrable (see section \ref{sec:int}).

\subsection{Motivation}
\la{mot}

To recall, the metric of $\AdS_3 \times \Sp^3$ space in global coordinates is\foot{We shall always use the string-frame metric and we mostly
omit the overall factor of string tension.}
\begin{equation} \label{1}
\extder s^2 = -(1+\rho^2) \extder t^2 + \frac{\extder \rho^2}{1+\rho^2} + \rho^2 \extder \psi^2 + (1-r^2) \extder \varphi^2 + \frac{\extder r^2}{1-r^2} + r^2 \extder \phi^2~.
\end{equation}
Its product with $\To^4$ can be embedded into 10d type IIB supergravity
by adding the NSNS 3-form flux
\begin{equation}
\label{2}
H_3 = \extder \hat{B} ~, \ \ \ \qquad \hat{B} = \rho^2 \, \extder t \wedge \extder \psi + r^2 \, \extder \varphi \wedge \extder \phi~.
\end{equation}
This geometry arises as the near horizon limit of the F1-NS5 solution (with equal charges).
The same metric can also be supported by the RR 3-form flux $F_3$ (corresponding to the near-horizon limit of the D1-D5 solution).
In view of the $\grp{SL}(2,\mathbb{R})$ symmetry of the type IIB equations one can also consider a
1-parameter ($|q| \leq 1$) family of mixed flux backgrounds (with constant dilaton $\P=\P_0$)
\begin{equation}
\label{3}
H_3 = \sqrt{1-q^2} \extder \hat{B}~, \qquad \qquad F_3 = e^{-\Phi_0} q \, \extder \hat{B}~ .
\end{equation}
Three more parameters can be added by applying
special $\grp{O}(d,d)$ or TsT transformations in the 4-torus directions.\foot{We shall refer to all transformations of the form ``T-duality - $\grp{GL}(d)$ coordinate redefinition - T-duality'' as TsT transformations. This includes both the usual TsT transformations where s stands for shift and the TrT transformations of \cite{OhlssonSax:2018hgc} where r stands for rotation.}
The resulting four-parameter family of supergravity backgrounds preserves 16 supersymmetries. 
The corresponding GS superstring model is integrable \ci{Babichenko:2009dk,Cagnazzo:2012se,Wulff:2014kja}.

Our aim will be to find a more general class of type IIB backgrounds
that preserve half of the maximal 16 supersymmetries and are still integrable.
Furthermore, we will find that they have regular curvature,
homogeneous fluxes\footnote{By homogeneous flux we mean that the corresponding tensor has
constant tangent space components and thus automatically satisfies some of the field equations (namely the ones analogous to the Maxwell equations).}
and constant dilaton and RR scalar. 

To motivate their construction let us first review the symmetries of the $\AdS_3 \times \Sp^3$ background.
The isometry algebra of $\AdS_3$ is $\alg{su}(1,1)_L \oplus \alg{su}(1,1)_R$, while that of $\Sp^3$ is $\alg{su}(2)_L \oplus \alg{su}(2)_R$.\foot{We use the labels $L$ and $R$ to distinguish the two copies of the algebra.}
Including fermions this is promoted to the superisometry algebra $\alg{psu}(1,1|2)_L \oplus \alg{psu}(1,1|2)_R$.
Four of the Killing spinors are associated to $\alg{su}(1,1)_L \oplus \alg{su}(2)_L$ and the other four are associated to $\alg{su}(1,1)_R \oplus \alg{su}(2)_R$.
When embedded into 10d supergravity with the $\To^4$ factor
the supersymmetries are doubled and the corresponding
GS superstring sigma model has 16 supersymmetries.

It is then natural to expect that backgrounds with 8 supersymmetries can be obtained by deforming either $\alg{psu}(1,1|2)_L$ or $\alg{psu}(1,1|2)_R$ in the superisometry algebra, while preserving the other copy.
One way to achieve this, which has the additional advantage of preserving integrability, is to apply a TsT transformation in the abelian isometries of one copy of $\alg{psu}(1,1|2)$.
The Cartan subalgebra is $\alg{u}(1)_L \oplus \alg{u}(1)_R$ for $\AdS_3$ and $\alg{u}(1)_L \oplus \alg{u}(1)_R$ for $\Sp^3$.
Without loss of generality, we may apply the TsT transformation in the isometries associated to the left copy $\alg{u}(1)_L \oplus \alg{u}(1)_L$.

Let us start with the background \eqref{1}-\eqref{3} with pure RR 3-form flux ($q=1$)
and with $x_r$, $r=1,\dots,4$, coordinates on $\To^4$.
The TsT transformation in the left Cartan directions (i.e.~in the combinations $t+\psi$ and $\varphi + \phi$ of coordinates in \rf{1})
with parameter $\k$ then produces the ``warped''
$\AdS_{3\k}$
and ``squashed''
$\Sp^3_\k$ metrics
\begin{equation} \label{4} \begin{aligned}
\extder s^2&= -(1+\rho^2) \extder t^2 + \frac{\extder \rho^2}{1+\rho^2} + \rho^2 \extder \psi^2 - \kappa^2 \big((1+\rho^2) \extder t - \rho^2\extder \psi\big)^2 \\
&\qquad + (1-r^2) \extder \varphi^2 + \frac{\extder r^2}{1-r^2} + r^2 \extder \phi^2 + \kappa^2 \big((1-r^2) \extder \varphi + r^2\extder \phi\big)^2 + \extder x_r \extder x_r~,
\end{aligned}
\end{equation}
and the following 3-form fluxes (with the dilaton remaining constant)
\begin{equation} \la{5}
H_3 = \kappa \sqrt{1+\kappa^2} \extder \CC ~, \qquad\qquad F_3 = e^{- \Phi_0} \sqrt{1+\kappa^2} \extder \hat{B}~.
\end{equation}
Here $ \hat{B}$ is as defined in \eqref{2} (i.e.~is given by the sum of AdS$_3$ and S$^3$ parts) while $\CC$
has instead a product structure ``mixing'' $\AdS_3$ and $\Sp^3$ coordinates
\begin{equation}
\label{6}
{\CC}=\left[ (1+\rho^2) \extder t - \rho^2 \extder \psi \right] \wedge \left[(1-r^2) \extder \varphi + r^2 \extder \phi \right]~.
\end{equation}
This background preserves the $\alg{psu}(1,1|2)_R$ superisometries.\foot{A similar TsT transformation in the right Cartan directions, which would preserve $\alg{psu}(1,1|2)_L$, gives rise to the same background
with
$t \to -t$ and $\varphi \to -\varphi.$}

It is useful to write this background in a different coordinate system where it has the form of a deformation
of the Hopf fibrations of $\AdS_3$ (over euclidean AdS$_2$ or 2d hyperbolic space $\Hy^2$) and $\Sp^3$ (over S$^2$).
Introducing the coordinates $(\zeta_1,\zeta_2,\bs)$ for AdS$_3$ and $( \x_1, \x_2, \s)$ for S$^3$ as
\begin{equation} \begin{aligned}\la{7}
t &= \tfrac{\zeta_1 - \zeta_2}{2}~, &\qquad \psi &= \tfrac{\zeta_1+\zeta_2}{2}~, &\qquad \rho &= \sinh \tfrac{\bs}{2}~, \\
\varphi &= \tfrac{\xi_1-\xi_2}{2}~, &\qquad \phi &= \tfrac{\xi_1+\xi_2}{2}~, &\qquad r &= \sin \tfrac{\s}{2}~,
\end{aligned}
\end{equation}
the deformed metric \rf{4} takes the form
\begin{equation} \label{8}
\begin{aligned}
\extder s^2 &= \tfrac{1}{4} \big( \sinh^2 \bs \, \extder \zeta_2^2 + \extder \bs^2 - (1+\kappa^2) \left(\extder \zeta_1 - \cosh \bs\, \extder \zeta_2\right)^2 \big)
\\
&\qquad +\tfrac{1}{4} \big( \sin^2 \s \, \extder \xi_2^2 + \extder \s^2 + (1+\kappa^2) \left(\extder \xi_1 - \cos \s\, \extder \xi_2\right)^2 \big) + \extder x_r \extder x_r ~.
\end{aligned}
\end{equation}
Writing the metric as a fibration over $\Hy^2 \times \Sp^2$ we see that the deformation simply rescales the fibres by $1+\k^2$.\foot{In the terminology of \cite{Chow:2009km} this deformation of $\AdS_3$ is also known as ``time-like squashed'' $\AdS_3$.}
The auxiliary 2-forms in \eqref{2} and \eqref{6} and their corresponding field strengths are then
\begin{equation}
\begin{aligned}
\hat{B} &=\tfrac{1}{2} \big( \sinh^2 \tfrac{\bs}{2} \, \extder \zeta_1 \wedge \extder \zeta_2+ \sin^2 \tfrac{\s}{2} \, \extder \xi_1 \wedge \extder \xi_2\big)~, \\
{\CC} &= \tfrac{1}{4} (\extder \zeta_1 - \cosh \bs \extder \zeta_2 ) \wedge (\extder \xi_1 - \cos \s \extder \xi_2)~,\la{99} \\
\extder \hat{B} &= \tfrac{1}{4} \big[ \sinh \bs \, \extder \zeta_1 \wedge \extder \zeta_2 \wedge \extder \bs + \sin \s \, \extder \xi_1 \wedge \extder \xi_2 \wedge \extder \s\, \big]~, \\
\extder {\CC} &= \tfrac{1}{4} \big[ \sinh \bs\, \extder \zeta_2 \wedge \extder \bs\, \wedge (\extder \xi_1 - \cos \s\, \extder \xi_2) + \sin \s (\extder \zeta_1 - \cosh \bs\, \extder \zeta_2) \wedge \extder \xi_2 \wedge \extder \s \big]~.
\end{aligned}
\end{equation}
Introducing the 1-form basis
\begin{equation}
\begin{aligned}
e^0 &= \tfrac{1}{2} \sqrt{1+\kappa^2}\left(\extder \zeta_1 - \cosh \bs \, \extder \zeta_2 \right)~,\qquad
e^1 =\tfrac{1}{2} \sinh \bs \, \extder \zeta_2~,\qquad
e^2 =\tfrac{1}{2}\extder \bs~, \\
e^3 &= \tfrac{1}{2} \sqrt{1+\kappa^2} \left( \extder \xi_1 - \cos \s\, \extder \xi_2 \right)~, \qquad \ \ e^4 =\tfrac{1}{2} \sin \s\, \extder \xi_2~, \qquad\ \ e^5 =\tfrac{1}{2}\extder \s~, \la{211}
\end{aligned}
\end{equation}
we can write the metric \rf{8} and the homogeneous 3-form fluxes \rf{99} as
\begin{align}
& \qquad \extder s^2 = - (e^0)^2 + (e^1)^2+ (e^2)^2+ (e^3)^2+ (e^4)^2 + (e^5)^2, \la{212} \\
\extder \hat{B} &= \tfrac{2}{\sqrt{1+\kappa^2}} \left( e^0 \wedge e^1 \wedge e^2 + e^3 \wedge e^4 \wedge e^5 \right)~, \qquad
\extder {\CC} = \tfrac{2}{\sqrt{1+\kappa^2}} \left(e^1 \wedge e^2 \wedge e^3 + e^0 \wedge e^4 \wedge e^5 \right)\ . \no
\end{align}

More general type IIB supergravity solutions can be obtained by applying
additional U-duality transformations (including $\grp{SL}(2,\Real)$ S-duality
and TsT in the 4-torus directions).
In the next section \ref{sols} we will show how
a more general family of flux backgrounds supporting the same deformed metric \rf{8} can be constructed directly as type IIB supergravity solution.
\iffa
Since S-duality transformations
need not ``commute'' with the integrability of the corresponding
superstring sigma model we will need an independent argument
for the integrability of this family of backgrounds, which will follow from the YB deformed supercoset model \cite{Delduc:2018xug} in section \ref{sec:int}.
\fi

\subsection{Type IIB supergravity solutions with \texorpdfstring{\adsk}{AdS3k x S3k} metric}
\label{sols}

The bosonic part of the type IIB supergravity action may be written as
\begin{equation}
\label{10}
\begin{aligned}
S_{10}=&
\int d^{10}x\sqrt{-G}\Big( e^{-2\Phi}\big(R+4\partial_\mu\Phi\partial^\mu\Phi-\tfrac{1}{2}|{{H}_3}|^2\big)
-\tfrac{1}{2} |{{F}_1}|^2 -\ha |{{F}_3}|^2 - \tfrac{1}{ 4} |{{F}_5}|^2 \Big)
\\ & \qquad \qquad \qquad \qquad+ \tfrac12 \int F_5\wedge F_3 \wedge B_2~,
\end{aligned}
\end{equation}
where
$H_3 = \extder B_2$ and $F_1$, $F_3$, $F_5$ are the RR field strengths,
\begin{equation} \label{eq:FC}
F_{n} = \extder C_{n-1} + H_3 \wedge C_{n-3}~,
\end{equation}
where $C_0, C_2,C_4$ are the RR potentials.
We use the notation
$|F_n|^2 = \tfrac{1}{n!} F_{\mu_1 \dots \mu_n} F^{\mu_1 \dots \mu_n}$.\foot{As usual, we relax the self-duality constraint on $F_5$, which is to be imposed at the level of equations of motion.}

We will consider homogeneous
flux backgrounds (meaning that their covariant derivatives vanish) and
also assume that the dilaton and the RR scalar have constant values.
Then the corresponding 10d supergravity equations simplify to\foot{Whenever the indices are written explicitly we remove the form degree index.}
\begin{equation} \begin{aligned}\la{12}
& R - \tfrac{1}{2} |{H_3}|^2 = 0 ~, \\
& R_{\mu \nu} - \tfrac{1}{4} H_{\mu \rho \sigma} H_\nu {}^{\rho \sigma} - \tfrac{1}{4} e^{2 \Phi_0} \Big( F_{\mu \rho \sigma} F_\nu{}^{\rho \sigma} + \tfrac{1}{4!} F_{\mu \rho \sigma \tau \upsilon} F_\nu^{\rho \sigma \tau \upsilon}- G_{\mu \nu } |{F_3}|^2\Big) = 0 ~, \\
& F_{\mu \nu \rho \sigma \tau} H^{\rho \sigma \tau} = 0~, \qquad F_{\mu \nu \rho \sigma \tau} F^{\rho \sigma \tau} = 0 ~,\qquad F_{\mu\nu\lambda} H^{\mu\nu\lambda} = 0 ~.
\end{aligned}
\end{equation}
Imposing the self-duality of the RR 5-form
gives $|{F_5}|^2=0$ and then the trace of \rf{12} implies that $|{H_3}|^2 = e^{2 \Phi_0} |{F_3}|^2$.

Now let us assume that the 10d metric and $F_5$ are of the $\Man^6 \times {\To}^4$ factorized form
\begin{align}
\extder s^2 & = G_{\mu \nu} \extder X^\mu \extder X^\nu = g_{mn}(x^k)\, \extder x^m \extder x^n + e^{A(x^k)} \extder x_r \extder x_r~,
\qquad
F_5 = \sum_{i=1}^3 F_3^{(i)} \wedge J_2^{(i)}~,\la{14}
\\
J_2^{(1)} & = \extder x_6 \wedge \extder x_7 - \extder x_8 \wedge \extder x_9, \quad J_2^{(2)} = \extder x_6 \wedge \extder x_8 + \extder x_7 \wedge \extder x_9, \quad J_2^{(3)} = \extder x_6 \wedge \extder x_9 - \extder x_7 \wedge \extder x_8, \no
\end{align}
where $m,n,k=0, \dots, 5$ and $r=6,7,8,9$ and we have defined the three orthogonal self-dual 2-forms $J_2^{(i)}$ on the torus $\To^4$.
We shall assume that the five 3-forms $H_3$, $F_3$ and $F_3^{(i)}$ have only $\Man^6$ components.
The 10d self-duality of $F_5$ implies that $F_3^{(i)}$ are self-dual on $\Man^6$ and hence $|{F_3}^{(i)}|^2=0$.

If we relax the self-duality condition on $F_3^{(i)}$, then
the bosonic part of the 6d supergravity action, corresponding to \rf{10} upon dimensionally reducing
on the 4-torus reads (see, e.g., \cite{Lavrinenko:1998hf})
\begin{equation}
\label{15}
\begin{aligned}
S &= \int \extder^6 x \sqrt{-g} \Big( e^{-2 (\Phi-A)} \big(R+4 (\partial \Phi)^2 - 4 \nabla^2 A - 5 (\partial A)^2 - \tfrac{1}{2} |H_3|^2\big) \\
&\hspace{178pt}- \tfrac{1}{2}e^{2 A} |F_3 |^2 - \tfrac{1}{2}e^{2 A} \sum_{i=1}^3 |F_3^{(i)}|^2 \Big)~.
\end{aligned}
\end{equation}
The variation over $\Phi$ and $A$ and setting the scalar $A=0$ implies
\begin{equation}\la{16}
|{F_3}|^2 + \sum_{i=1}^3 |F_3^{(i)}|^2 =0~,
\end{equation}
which, of course, also follows directly from the 10d equations in \rf{12}.
Together with the self-duality constraints we are then left with the following equations
\begin{align} \label{17}
R = |{H_3}|^2=|{F_3}|^2=|{F_3^{(i)}}|^2= H_3 \cdot F_3=H_3 \cdot F_3^{(i)}=F_3 \cdot F_3^{(i)} = 0~, \\ \la{18}
R_{mn} - \tfrac{1}{4} H_{mkl} H_n {}^{kl} - \tfrac{1}{4} e^{2 \Phi_0} \Big( F_{mkl} F_n{}^{kl} + \sum_{i=1}^3 (F^{(i)})_{mkl} (F^{(i)})_n{}^{kl}\Big) = 0~.
\end{align}
For constant scalars the five 3-forms $H_3$, $F_3$, $F^{(i)}_3$ enter the 6d action \rf{15} on an equal footing,
i.e.~there is an $SO(5)$ symmetry relating them, which is implied by U-duality.

Let us now take the $\Man^6$ metric to be given by the deformed ${\rm AdS}_{3\k} \times {\rm S}^3_\k$ metric in \rf{4} or \rf{8}, which
has the vanishing 6d Ricci scalar as required by \rf{17}.
The basic 3-forms $ \extder \hat{B}$ and $\extder {\CC}$ in \rf{2} and \rf{6}
that enter the simplest example of the background supporting \adsk
satisfy
\be \la{19} |\extder \hat{B}|^2 = |\extder {\CC}|^2 = \extder \hat{B} \cdot \extder {\CC}=0 ~ . \ee
This implies that if we take the five 3-forms $H_3, F_3, F^{(i)}_3$
to be given by linear combinations of $\extder \hat{B}$ and $\extder {\CC}$ in \rf{212} as
\begin{align} & \bF_3\equiv \big(H_3, F_3, F^{(1)}_3, F^{(2)}_3,F^{(3)}_3 \big) = \textbf{z}_1\, \extder \hat{B}+ \textbf{z}_2\, \extder {\CC}\ , \la{20}
\end{align}
where
$\textbf{z}_1$ and $\textbf{z}_2$ are constant 5-vectors,
then all the equations in \eqref{17} will be automatically satisfied.
Explicitly, we shall use the following ansatz\foot{In what follows
we will set $\Phi_0 =0$ (the dependence on the constant factors
$e^{-\Phi_0}$ in the RR fluxes can easily be restored).
Let us also mention that $F_3= d C_2$ (we set $C_0=0$) and that $F_5$ may be written as $F_5 = \extder C'_4$ since
the additional term in \eqref{eq:FC}
$H_3 \wedge C_2 \sim \tfrac{1}{2} \extder (\hat{B} \wedge \hat{B}) + \tfrac{1}{2} \extder (\check{B} \wedge \check{B})$
is a total derivative.
We will always discard total derivative terms in the NSNS and RR potentials.}
\begin{align}
H_3 &= s_1 \extder \hat{B}+ s_2 \extder {\CC} ~, \qquad
F_3 =
y_1 \extder \hat{B}+ y_2 \extder {\CC} ~, \label{22}\\
F_5 &=
(y_3 \extder \hat{B} + y_4 \extder {\CC} ) \wedge J_2^{(1)} + (y_5 \extder \hat{B} + y_6 \extder {\CC} ) \wedge J_2^{(2)}+(y_7 \extder \hat{B} + y_8 \extder {\CC} ) \wedge J_2^{(3)} ~,\no
\end{align}
where $s_1,s_2$ and $y_1, \dots, y_8$ are ten real parameters, which are the components of $\textbf{z}_1$ and $\textbf{z}_2$
in \rf{20}
\begin{equation}\la{23}
\textbf{z}_1 = (s_1,y_1,y_3,y_5,y_7)~,\qquad \textbf{z}_2=(s_2,y_2,y_4,y_6,y_8)~.
\end{equation}
Starting with the \adsk metric \rf{8} and \rf{212} and the fluxes in \rf{20} and \rf{22}
we conclude that the supergravity equations \rf{12} or \rf{17} and \rf{18} are satisfied provided that
the ten constants in \rf{23} are subject to the following constraints
\begin{equation} \label{24}
\begin{aligned}
\textbf{z}_1 \cdot \textbf{z}_2 = 0~, \qquad \lVert\textbf{z}_1\rVert^2= 1+\kappa^2 ~, \qquad
\lVert\textbf{z}_2\rVert^2= \kappa^2 (1+\kappa^2) ~.
\end{aligned}
\end{equation}
The special background in \rf{5} is obviously a particular solution of \rf{22} and \rf{24} with (for $\Phi_0=0$)
\be\la{244}
\textbf{z}_1 = (0, \, \sqrt{1 + \k^2},\, 0,0,0)~,\qquad \textbf{z}_2=(\k \sqrt{1 + \k^2} ,\, 0,0,0,0)
\ .
\end{equation}
Let us now discuss the meaning and consequences of the constraints \rf{24}.

\subsubsection{U-duality transformations}
\la{udu}
The U-duality group
of 6d maximal supergravity is $\grp{Spin}(5,5)$ with
maximal compact subgroup $\grp{Spin}(5)$, which is locally isomorphic to $\grp{SO}(5)$.
Correspondingly, the equations \eqref{24} are invariant under simultaneously rotating the two vectors $\textbf{z}_1$ and $\textbf{z}_2$ by $\mathcal R \in \grp{SO}(5)$.
This has a natural geometric interpretation in terms of T-dualities and rotations in the 4-torus directions (or TsT transformations), and S-duality rotations.

An example of a
TsT transformation involving a pair of coordinates $x_r$ and $x_s$ of the torus is to first apply T-duality $x_r\to \td x_r$, then rotate (with parameter $\beta$)
\begin{equation}\la{25}
\td x_r \rightarrow \td x_r \cos \beta - x_s \sin \beta~, \qquad x_s \rightarrow x_s \cos \beta + \td x_r \sin \beta~,
\end{equation}
and finally T-dualise back $\td x_r \to x_r$.
It is easy to see that TsT
in $x_6$ and $x_7$ (or $x_8$ and $x_9$)
results in a rotation in the $(y_1,y_3)$ and $(y_2, y_4)$ planes of the 10-parameter space in \rf{23}.
Similarly, TsT in $x_6$ and $x_8$ (or $x_7$ and $x_9$) results in a rotation in the $(y_1,y_5)$ and $(y_2, y_6)$ planes, while TsT in $x_6$ and $x_9$ (or $x_7$ and $x_8$) gives a rotation in the $(y_1,y_7)$ and $(y_2, y_8)$ planes.

S-duality transformations (with $\P_0=0$ and $C_0=0$) leave the string-frame metric invariant and just rotate the NSNS
$H_3$ and RR $F_3$ forms into each other so that their coefficients in \rf{20} and \rf{22}
change as
\begin{equation} \label{26} \begin{aligned}
s_1
&= s_1 \cos \alpha + y_1 \sin \alpha ~, &\qquad s_2
&=s_2 \cos \alpha + y_2 \sin \alpha~, \\
y_1
&= y_1 \cos \alpha - s_1 \sin \alpha~, &\qquad y_2
&= y_2 \cos \alpha - s_2 \sin \alpha~.
\end{aligned}
\end{equation}
This is a simultaneous rotation in the $(s_1,y_1)$ and $(s_2,y_2)$ planes of the parameter space with angle $\alpha$.
More general $\grp{SO}(5)$ rotations are then
obtained by combining the above TsT and S-duality transformations.

\subsubsection{Seed solutions}
\la{seed}

In the simplest case of $\k=0$ (corresponding to undeformed \adss metric)
the third equation in \eqref{24} implies that $\textbf{z}_2=0$, which means that
the fluxes in \rf{22} do not depend on d$\CC$.
The coefficients of d$\hat B$ then satisfy
$\lVert\textbf{z}_1\rVert=1$ such that there are four independent parameters (parametrising a 4-sphere). Equivalently,
the general solution is in one to one correspondence with rotations $\RR \in \grp{SO}(5) /\grp{SO}(4)$
\begin{equation}\la{27}
\textbf{z}_1 = \RR \textbf{v}_1~,
\end{equation}
where $\textbf{v}_1$ is a fixed unit 5-vector that is invariant under an $\grp{SO}(4)$ subgroup of $\grp{SO}(5)$.
We can choose $\textbf{v}_1$ to represent a simple ``seed'' solution, corresponding, e.g., to pure NSNS or pure RR flux.
The most general background (with four parameters)
is then obtained by applying the rotation $\RR$, which can be decomposed into an
S-duality rotation (producing the mixed flux background) and three additional TsT transformations in the torus directions as described in section \ref{udu}.

For $\kappa \neq 0$ we have ten parameters
obeying three equations in \rf{24}, thus leaving seven free parameters.
The space of solutions is now in one to one correspondence with rotations $\RR \in \grp{SO}(5)/\grp{SO}(3)$ and the most general solution to \eqref{24} can be parametrised as
\begin{equation}\la{311}
\textbf{z}_1 = \sqrt{1+\kappa^2} \RR \textbf{v}_1~, \qquad\qquad \textbf{z}_2 = \kappa \sqrt{1+\kappa^2} \RR \textbf{v}_2 ~,
\end{equation}
where $\textbf{v}_1$ and $\textbf{v}_2$ are two orthogonal unit-norm 5-vectors, which are invariant under an $\grp{SO}(3)$ subgroup of $\grp{SO}(5)$.
Note that the orthogonality constraint forbids pure NSNS solutions for $\kappa \neq 0$.\foot{In the pure NSNS case we have
$y_k=0$ in \rf{22}, but then $\textbf{z}_1 \cdot \textbf{z}_2=0$ in \rf{24} implies $s_1 s_2=0$ which is inconsistent with the remaining two equations if $\k\not=0$ and $\k\not=i$.}

Examples of simple seed solutions include (we indicate only the non-zero fluxes)
\begin{enumerate}[(i)]
\item $H_3\not=0$, $F_3\not=0$ (e.g., the background \rf{4} and \rf{5} or \rf{244});
\item $F_3\not=0$, $F_5\not=0$;
\item $H_3\not=0$, $F_5\not=0$ (related to (i) by double T-duality);
\item $F_5\not=0$ (related to (ii) by double T-duality).
\end{enumerate}
Additional parameters may be turned on through combinations of S-duality and TsT transformations in the torus directions. For instance,
starting from case (i) with $\textbf{v}_1 = (1,0,0,0,0)$ and $\textbf{v}_2 = (0,1,0,0,0)$ we can do
three TsT transformations in the torus directions; this
leaves $\textbf{z}_1=\textbf{v}_1$ invariant, while introducing three new parameters in $\textbf{z}_2$.
Then using combinations of S-duality and TsT transformations we can
add four more parameters, thereby generating the full seven-parameter family of solutions.

As discussed in section \ref{mot}, the deformation parameter $\k$ in the metric \rf{4} or \rf{8} can be turned
on by starting with the undeformed \adss metric \rf{1} and applying a TsT transformation
in the left Cartan directions ($\zeta_1$ and $\xi_1$).
Therefore, starting with the seed solution of \rf{24} with $\k=0$ and pure RR 3-form flux
the full seven-parameter family of solutions can be obtained by first turning on
$\k$ using a TsT transformation in the left Cartan directions.

\subsubsection{TsT in left Cartan directions}
\la{cart}

To complete our discussion of symmetry transformations and the constraints \rf{24}
let us give details of the TsT transformation in the left Cartan directions.

We shall use the Hopf fibration parametrisation \eqref{7} and consider the following TsT transformation in $\zeta_1$ and $\xi_1$ with a parameter $\gamma$
\be \la{29} {\rm T}: \ \xi_1\to \td \xi_1 ~, \qquad
\zeta_1 \rightarrow \zeta_1+ \gamma \td \xi_1 ~, \qquad {\rm T}: \ \td \xi_1\to \td {\td {\xi}}_1\equiv \x_1 ~. \ee
Up to a coordinate redefinition and a total derivative $B$-field, this takes the metric \rf{8} and fluxes \rf{22}
with parameters $(\kappa,s_a,y_k)$
to the same metric and fluxes with new parameters $(\hat\kappa,\hat s_a,\hat y_k)$
where
\begin{align}\label{30}
\hat{s}_1 &= s_1~, &\qquad \hat{s}_2 &= \frac{4 (\hat{\kappa}^2-\kappa^2)- \gamma s_2 (1+\hat{\kappa}^2)}{\gamma(1+\kappa^2)}~, \\
\hat{y}_k &= \frac{ 4y_k - \gamma s_1 y_{k+1}}{\sqrt{(4-\g s_2)^2-\g^2 (1+\kappa^2)^2}} ~, &\qquad \hat{y}_{k+1} &= \frac{4y_{k+1} - \gamma s_1 y_k}{\sqrt{(4-\g s_2)^2-\g^2 (1+\kappa^2)^2}}~, \qquad k=1,3,5,7~,\no
\end{align}
and $\g=\g(\k, \hat \k, s_1,s_2)$ is given by
\begin{equation} \la{31}
\gamma = \frac{-4s_2 (1+\hat{\kappa}^2) + 4\sqrt{1+\kappa^2} \sqrt{(1+\kappa^2)(1+\hat{\kappa}^2) (\hat{\kappa}^2-\kappa^2) - s_1^2 (\hat{\kappa}^2-\kappa^2 ) + s_2^2 (1+\hat{\kappa}^2)}}{ (1+\kappa^2)^2 (1+\hat{\kappa}^2)-s_1^2 (1+\kappa^2) - s_2^2 (1+\hat{\kappa}^2)}~.
\end{equation}
Note that, since TsT transformations map supergravity solutions into supergravity solutions, the constraints \eqref{24} are
still obeyed by the transformed coefficients $(\hat\kappa, \hat s_a, \hat y_k)$.

If we start with undeformed background ($\kappa=0,\, \lVert \textbf{z}_1 \rVert=1,\, \textbf{z}_2=0$), the TsT transformation in the left Cartan directions then gives a background with non-zero $\hat \k$.
The determinant
of the coordinate redefinition needed to bring the metric into the standard deformed
form \eqref{8} is
$D= \frac{(1+\hat{\kappa}^2)(1-s_1^2)}{1+\hat{\kappa}^2-s_1^2}. $
In the pure NSNS case we have $\textbf{z}_1= (\pm 1, 0, \dots, 0)$, i.e., $s_1^2=1$, which means that the determinant vanishes, hence it is not
a good starting point,\foot{The TsT transformation of the undeformed pure NSNS solution is again the undeformed pure NSNS solution.}
but all other cases are.\foot{Note that $\lVert \textbf{z}_1 \rVert =1$ imposes $s_1^2 \leq 1$ and if $\hat{\kappa}^2>0$ then the denominator in $D$ never vanishes.}
From \rf{30} and \rf{31} for $\kappa = 0$ we have that
\begin{equation} \begin{aligned}\la{33}
& \hat{s}_1 = s_1~, \qquad \hat{s}_2 = \hat{\kappa} \sqrt{1+\hat{\kappa}^2 - s_1^2}~,\qquad
\g=\frac{4 \hat{\kappa}}{\sqrt{1+\hat{\kappa}^2-s_1^2}} ~, \\
&\hat{y}_k = \frac{\sqrt{1+\hat{\kappa}^2-s_1^2}}{\sqrt{1-s_1^2}}y_k ~, \qquad
\hat{y}_{k+1} = - \frac{\hat{\kappa} s_1}{\sqrt{1-s_1^2}}y_k~, \qquad k=1,3,5,7 \ .
\end{aligned}
\end{equation}
The corresponding background has four independent parameters in addition to $\hat{\kappa}$, which can be taken to be $s_1$ and $y_k$ ($k=1,3,5,7$) subject to $y_1^2+y_3^2+y_5^2+y_7^2=1-s_1^2$.
Setting $\hat{s}_1 = \sqrt{1-q^2}$ and introducing an
auxiliary 4-vector $\hat{\textbf{u}} = (\hat{y}_1, \hat{y}_3, \hat{y}_5, \hat{y}_7)$
the resulting coefficients \rf{23} of the supergravity background can be written as
\begin{equation}
\label{34}
\hat{\textbf{z}}_1 = (\sqrt{1-q^2}, \hat{\textbf{u}})~, \qquad \hat{\textbf{z}}_2 = \Big(\hat{\kappa} \sqrt{q^2+\hat{\kappa}^2}, - \frac{\hat{\kappa} \sqrt{1-q^2}}{\sqrt{q^2+\hat{\kappa}^2}} \hat{\textbf{u}}\Big)~, \qquad \Vert\hat{\textbf{u}}\Vert^2=q^2+\hat{\kappa}^2 ~.
\end{equation}
This is the most general deformed background that can be obtained from the undeformed one
by TsT transformations alone.
Since the undeformed string sigma model is integrable, and TsT preserves
integrability, the same applies to this background as well.
In section \ref{sec:int} we will prove the classical integrability of the string sigma model for the full seven-parameter family of solutions.

An example of a solution corresponding to
\rf{34} is found by starting from the undeformed \adss background supported by mixed flux
\eqref{3} with one free parameter $|q| \leq 1$, i.e.~with
$\textbf{z}_1=(\sqrt{1-q^2},q,0,0,0)$ and $\textbf{z}_2=0$ (cf. \rf{22} and \rf{23}). Then, assuming $q \neq 0$,
the above TsT transformation gives
a deformed ${\rm AdS}_{{3\hat \k}} \times {\rm S}^3_{{\hat \k}} $ background with $\gamma = \frac{4\hat\ka}{\sqrt{q^2+\hat\k^2}}$ and
\begin{equation}
\la{35}
\hat{\textbf{z}}_1 = (\sqrt{1-q^2},\sqrt{q^2+\hat{\kappa}^2},0,0,0)~, \qquad \hat{\textbf{z}}_2 = (\hat{\kappa} \sqrt{q^2+\hat{\kappa}^2},-\hat{\kappa} \sqrt{1-q^2},0,0,0)~,
\end{equation}
i.e.~with only the $H_3$ and $F_3$ fluxes non-vanishing.

\section{Supersymmetry}
\label{sec:susy}

The undeformed $\rm AdS_3 \times S^3 \times T^4$ solution
preserves $\tfrac12$ of maximal 10d supersymmetry, i.e.~has 16 supercharges.
Let us now show that the family of deformed backgrounds \rf{4}, \rf{22}--\rf{24} preserves $\tfrac14$ of maximal supersymmetry, i.e.~admits 8 Killing spinors.

This may be at first surprising.
Indeed, in general, T-dualities in 4-torus directions and S-duality should preserve supersymmetry.
However, the T-duality along
Hopf fibres (which in the present case is responsible for introducing the deformation parameter
$\k$) may break supersymmetry of the supergravity background \ci{Duff:1998us,Duff:1998cr}.\foot{The full supersymmetry may still be ``hidden'' at the level of full string theory (cf. \ci{Duff:1998us}). This may be indeed related to the integrability
of the underlying supercoset model discussed in section \ref{sec:int}.}

In general, there are two 10d 
type IIB supergravity Killing spinor equations, associated with the invariance of the gravitino $\psi_\mu$
and the dilatino $\lambda$ fields under the
supersymmetry variations (see, e.g., \cite{Hassan:1999bv,Cvetic:2002nh,Papadopoulos:2003jk})\foot{Construction of Killing spinors
on AdS spaces and spheres was discussed, e.g., in \ci{Lu:1998nu,Duff:1998cr}.}
\begin{align}
\delta \psi_\mu&= D_\mu \ep =\big( \nabla_\mu + \tfrac{1}{8} H_{\mu a_1 a_2} \Gamma^{a_1 a_2} \sigma_3 + \mathcal S \Gamma_\mu \big) \ep =0 ~, \la{51}\\
\delta\lambda &= \big[\Gamma^\mu \partial_\mu \Phi +\tfrac{1}{12} H_{a_1 a_2 a_3} \Gamma^{a_1 a_2 a_3} \sigma_3 + e^{\Phi} \left(- i
\slashed{F_1}
\sigma_2 +\tfrac{1}{12} \slashed{F_3} \sigma_1 \right)\big] \ep =0 ~,\nn
\nabla_\mu & = \partial_\mu + \tfrac{1}{4} \omega_\mu^{a_1 a_2} \Gamma_{a_1 a_2}\ , \qquad
\mathcal S \equiv -\tfrac{1}{8} e^\Phi\left( i \slashed{F_1} \sigma_2 + \tfrac{1}{3!} \slashed{F_3} \sigma_1
+ i \tfrac{1}{2 \cdot 5!} \slashed{F_5} \sigma_2 \right)~.\la{52}
\end{align}
Here $\epsilon=(\epsilon^1, \epsilon^2)$ is a doublet of 32-component Majorana-Weyl spinors with $\sigma_k$ being Pauli matrices acting on $I=1,2$.
$\Gamma^a$ are $32 \times 32$ 10d Dirac matrices, $\anticom{\Gamma^a}{\Gamma^b} = 2 \eta^{ab}$ and $\slashed{F}_m \equiv F_{a_1 \dots a_{m}} \Gamma^{a_1 \dots a_m}$, with $\Gamma^{a_1 \dots a_m} = \Gamma^{a_1} \dots \Gamma^{a_m}$. We use greek letters for the spacetime indices and latin letters $a, a_j$ for tangent space indices. 

The background in \rf{22} that we are interested in
has constant dilaton and RR scalar (i.e.~$F_1=0$)
and 6d self-dual $H_3$ and $F_3$. More precisely, in the vielbein basis defined in \eqref{211}, the fluxes in \rf{22} are such that
\begin{equation} \begin{aligned}
\slashed{H}_3 &= \frac{2}{\sqrt{1+\kappa^2}} (s_1 \Gamma^{012} +s_2 \Gamma^{123})(1+\Gamma^{11} \Gamma^{6789})~, \\
\slashed{F}_3 &= \frac{2}{\sqrt{1+\kappa^2}} (y_1 \Gamma^{012} +y_2 \Gamma^{123})(1+\Gamma^{11} \Gamma^{6789})~, \\
\slashed{F}_5 &= \frac{2}{\sqrt{1+\kappa^2}}(\Gamma^{012} I +\Gamma^{123} J )(1+\Gamma^{11})(1+\Gamma^{6789})~,
\end{aligned}
\end{equation}
where $\Gamma^{11}= \Gamma^{0123456789}$, $I =y_3 \Gamma^{67} + y_5 \Gamma^{68} + y_7 \Gamma^{69}$ and $J=y_4 \Gamma^{67} + y_6 \Gamma^{68} + y_8 \Gamma^{69}$.

The equation $\delta \lambda =0$ is then trivially satisfied if we use the type IIB chirality condition
$\epsilon^I = \hat{\Gamma} \epsilon^I$, where $\hat{\Gamma} = \frac{1}{2} (1-\Gamma^{11})$, together with the condition $\epsilon^I = \mathcal P \epsilon^I$, where the projector $\mathcal P = \frac{1}{2} (1+\Gamma^{6789})$.\foot{In the
corresponding GS action the Killing spinors correspond to the fermionic isometries.
The fermionic coordinates $\vartheta$
are the negative chirality MW spinors:
$\Gamma^{11} \vartheta = - \vartheta$. The RR flux bispinor $\mathcal S$
enters the GS action as $\bar{\vartheta} \Gamma_\mu \mathcal S \Gamma_\nu \vartheta$ where
$\bar{\vartheta}= \vartheta^t \mathcal C$ and the charge conjugation matrix is $\mathcal C = i \sigma_2 \otimes 1_{16}$ so that $\bar{\vartheta} \Gamma^{11} = + \bar{\vartheta}$.
The Dirac matrices are $\Gamma^a = \begin{pmatrix} 0 & (\gamma^a)^{\alpha \beta} \\
(\gamma^a)_{\alpha \beta} & 0\end{pmatrix}$. We can then write $\bar{\vartheta} \Gamma_\mu \mathcal S \Gamma_\nu \vartheta = \bar{\vartheta} \Gamma \Gamma_\mu \hat{\Gamma}\mathcal S \Gamma \Gamma_\nu \hat{\Gamma} \vartheta$, with $\hat{\Gamma} = \frac{1}{2} (1- \Gamma^{11})$.} This leaves $2 \times 8 =16 $ independent components of $\ep$. It also ensures that the Killing spinors do not depend on the 4-torus directions $6,7,8,9$.

The equations $D_\m \ep=0$ with the index $\mu$ corresponding to the directions $(\zeta_1, \bs, \xi_1, \s)$ in \rf{8}
are first order differential equations of the form $\partial_\mu \epsilon = \Omega_\mu \epsilon$ with coefficients $\Omega_\mu$ that do not
explicitly depend on the coordinates, see Appendix \ref{susy}.
The compatibility conditions $\com{\Omega_{\zeta_1}}{\Omega_{\xi_1}} \epsilon=\com{\Omega_{\bs}}{\Omega_{\s}} \epsilon=0$ are immediately satisfied. The other compatibility conditions are satisfied in two cases.
One option (A) is to choose the parameters of the fluxes in \rf{23} so that
\begin{equation}
\label{eq:KScond1}
{\rm A}: \qquad \Vert \textbf{z}_1 \Vert^2= (1+\kappa^2)^2~, \qquad \textbf{z}_2=0~,
\end{equation}
and impose no condition on $\epsilon$.
The other option (B) is to impose the two equations
\begin{equation}
\label{eq:KScond2}
{\rm B}: \qquad\qquad \Vert \textbf{z}_1 \pm \textbf{z}_2 \Vert^2 = (1+\kappa^2)^2~,
\end{equation}
together with an additional constraint $\epsilon = \mathcal Q \epsilon$ (where the operator $\mathcal Q$ depends on the coefficients) which halves the number of independent Killing spinors.
In both cases the solution is then given by
\begin{equation}
\label{eq:KSchi}
\epsilon = \exp \left( \Omega_{\zeta_1} \zeta_1+\Omega_{\bs} \bs + \Omega_{\xi_1} \xi_1 + \Omega_\s \s \right) \chi~,
\end{equation}
where the spinor $\chi=\mathcal Q \chi$ ($\mathcal Q=1$ for case (A)) depends only on the coordinates $ \zeta_2$ and $\xi_2$.
Further using that
\begin{equation}
\com{\Omega_{\zeta_1}}{\Omega_{\zeta_2}} \epsilon= \com{\Omega_{\xi_1}}{\Omega_{\zeta_2}} \epsilon=\com{\Omega_{\zeta_1}}{\Omega_{\xi_2}} \epsilon= \com{\Omega_{\xi_1}}{\Omega_{\xi_2}} \epsilon= \com{\Omega_{\s}}{\Omega_{\zeta_2}} \epsilon=\com{\Omega_{\bs}}{\Omega_{\xi_2}} \epsilon=0~,
\end{equation}
the two remaining equations $D_{\zeta_2} \epsilon=0$ and $D_{\xi_2}\epsilon=0$ simplify to
\begin{equation} \label{eq:KSint}
\begin{aligned}
\partial_{\zeta_2} \chi &= \hat{\Omega}_{\zeta_2} \chi~, \qquad \hat{\Omega}_{\zeta_2}= \exp \left(- \Omega_{\bs} \bs\right) \Omega_{\zeta_2}\exp \left( \Omega_{\bs} \bs \right)~, \\
\partial_{\xi_2} \chi &= \hat{\Omega}_{\xi_2} \chi~, \qquad \hat{\Omega}_{\xi_2}=\exp \left(- \Omega_{\s} \s \right) \Omega_{\xi_2}\exp \left( \Omega_{\s} \s \right)~. \\
\end{aligned}
\end{equation}
Let us recall that $\chi$ depends on $\zeta_2$ and $\xi_2$, $\Omega_\sigma$ and $\Omega_\theta$ are constant, $\Omega_{\zeta_2}$ depends on $\sigma$ and $\Omega_{\xi_2}$ depends on $\theta$. Therefore, for \eqref{eq:KSint} to be satisfied, we need $\hat{\Omega}_{\zeta_2}$ and $\hat{\Omega}_{\xi_2}$ to be constants, i.e.~without $\sigma$ or $\theta$ dependence. This is indeed the case if one further imposes
\begin{equation}
\label{eq:KScond3}
\Vert \textbf{z}_1 \Vert^2=1+\kappa^2~.
\end{equation}
One then shows that $\com{\hat{\Omega}_{\zeta_2}}{\hat{\Omega}_{\xi_2}}\chi =0$ and therefore $\chi= \exp (\hat{\Omega}_{\zeta_2} \zeta_2 + \hat{\Omega}_{\xi_2} \xi_2 ) \chi_0$, with $\chi_0$ a constant vector obeying $\chi_0=\mathcal Q \chi_0$.

For option (A) we have $\mathcal Q =1$ and the conditions \eqref{eq:KScond1} and \eqref{eq:KScond3} together imply
$\kappa=0$. In this case the background is maximally supersymmetric in 6d (with 16 Killing spinors), and
corresponds to the undeformed $\AdS_3 \times \Sp^3 \times \To^4$.

To have $\kappa \neq 0$ one needs to consider the option B, with the additional projection $\epsilon = \mathcal Q \epsilon$.
The conditions \eqref{eq:KScond2} and \eqref{eq:KScond3} together are equivalent to the
constraints \eqref{24} following from the supergravity field equations.
The resulting background thus preserves only half of the original 16 supersymmetries, i.e.~admits 8 Killing spinors. 

\section{Classical integrability of the superstring sigma model}
\label{sec:int}

In section \ref{sec:2}, starting from the $\AdS_3 \times \Sp^3 \times \To^4$ background supported by mixed flux \rf{3}, we constructed the 8-parameter background (including $\ka$) with metric~\rf{8}, constant dilaton and axion, and fluxes~\rf{99}, \rf{22} subject to the constraints \rf{23}, \rf{24}.
This background preserves 8 supersymmetries (for $\ka \neq 0$) and can be generated by a combination of TsT transformations {\it and} S-dualities (i.e.~not TsT transformations alone).
While TsT transformations preserve the classical integrability of the corresponding string sigma model, a priori S-dualities do not.
Our aim in this section is to show that the full 8-parameter background defines an integrable string sigma model.

To do this we recall that in a particular $\varkappa$-symmetry gauge the $\AdS_3\times \Sp^3\times \To^4$ GS string sigma model with pure RR flux can be written as a semi-symmetric space sigma model plus 4 compact bosons \cite{Rahmfeld:1998zn}.
The relevant $\Integer_4$ supercoset is
\begin{equation}\la{400}
\frac{\grp{PSU}(1,1|2) \times \grp{PSU}(1,1|2)}{\grp{SU}(1,1) \times \grp{SU}(2)} ~,
\end{equation}
where the isotropy group $\grp{SU}(1,1) \times \grp{SU}(2)$ is the diagonal bosonic subgroup.
This model is known to be classically integrable \cite{Babichenko:2009dk}.
It is also known that it admits integrable deformations known as YB deformations
(see \cite{Hoare:2021dix} and references there).

In general, the YB deformation \cite{Klimcik:2002zj,Delduc:2013qra} depends on an antisymmetric linear operator $R$, known as an R-matrix, acting on the symmetry algebra $\alg{g}$ of the model.
This operator solves the (modified) classical Yang-Baxter equation ((m)cYBe)
\begin{equation}\la{401}
[RX,RY] - R[X,RY] - R[RX,Y] + c^2[X,Y] =0~, \qquad X,Y\in \alg{g} ~.
\end{equation}
When this R-matrix satisfies a so-called unimodularity condition, that is when the trace of the structure constants of the dual Lie algebra $\alg{g}_R$ with Lie bracket $[X,Y]_R = [X,RY] + [RX,Y]$ vanishes, the deformed string sigma model remains Weyl invariant, i.e.~the background fields solve the
supergravity equations \cite{Borsato:2016ose} (see also \cite{Alvarez:1994np,Elitzur:1994ri}).

The case of interest for us is $c = i$, i.e.~the non-split mcYBe.
For the supergroups relevant for constructing string sigma models, e.g.~$\grp{PSU}(2,2|4)$ for $\AdS_5 \times \Sp^5$ and $\grp{PSU}(1,1|2)$ for $\AdS_3 \times \Sp^3$ and $\AdS_2\times\Sp^2$ backgrounds, there exist unimodular solutions to the mcYBe \cite{Hoare:2018ngg,Seibold:2019dvf}.
Introducing Cartan generators $H_i$ and positive and negative roots $E_\alpha$ and $F_\alpha$, these are the so-called Drinfel'd-Jimbo (DJ) solutions ($R(H_i) = 0$, $R(E_\alpha) = i E_\alpha$, $R(F_\alpha) = -i F_\alpha$) built from a Cartan-Weyl basis with all fermionic simple roots.
The existence of these solutions is crucial for constructing YB deformations based on solutions to the non-split mcYBe that define Weyl invariant string sigma models \ci{Hoare:2018ngg}.

Other Drinfel'd-Jimbo solutions exist based on different Dynkin diagrams that do not satisfy the unimodularity condition and are not Weyl invariant \cite{Arutyunov:2015qva,Arutyunov:2015mqj,Hoare:2018ngg,Seibold:2019dvf}.
Nevertheless, one can still fix a light-cone gauge and it seems reasonable to expect that the resulting models are quantum integrable.
Indeed, for e.g.~$\AdS_5 \times \Sp^5$, it is possible to conjecture an exact S-matrix for the corresponding 8+8 transverse degrees of freedom \cite{Beisert:2008tw,Hoare:2011wr,Arutyunov:2013ega,Seibold:2020ywq}.
The different S-matrices coming from different Dynkin diagrams are related by fermionic twists \cite{Seibold:2020ywq} and share the same Bethe equations \cite{Seibold:2021rml}.
However, this does not necessarily imply that the string sigma models are fully equivalent.
Indeed, the lack of Weyl invariance suggests that there may be issues with light-cone gauge-fixing beyond the classical level, e.g., leading to an anomaly in the full deformed global symmetry. Here we will always consider unimodular R-matrices so that the corresponding sigma model is Weyl invariant. An example of deformation based on a non-unimodular R-matrix is provided in Appendix \ref{app:dist}.

For the $\AdS_3 \times \Sp^3 \times \To^4$ background, the direct product structure of the superisometry group means that the model admits a richer space of deformations.
In particular, an additional WZ term can be included in the coset sigma model corresponding to the mixed flux background \cite{Cagnazzo:2012se}, and the two copies of $\grp{PSU}(1,1|2)$ can be deformed with different strengths \cite{Hoare:2014oua}.
While the full space of YB deformations has not been fully understood, a three-parameter model is known based on deforming the mixed flux model with the DJ R-matrix on each copy of $\grp{PSU}(1,1|2)$.
This generalises the bi-YB deformation of the PCM plus WZ term \cite{Delduc:2017fib,Klimcik:2019kkf} (see also \cite{Klimcik:2008eq,Delduc:2014uaa}) to $\Integer_4$ supercosets.
The action of this three-parameter model is \cite{Delduc:2018xug}
\begin{align}
S_{\eta_L,\eta_R,k} &= T \int \extder^2 x \,\STr \Big(J_+ \big( (1-(2 P_2 + P_F) k W \Omega_-) d_- + P_F k W (1-d_-) \big) \frac{1}{1+\Omega_- d_-} J_- \Big) \no \\
&\qquad - 4 T k \int \extder^3 x \, \epsilon^{\mu \nu \rho} \STr \left( \tfrac{2}{3} W J_\mu^{(2)} J_\nu^{(2)} J_\rho^{(2)} + W \com{J_\mu^{(1)}}{J_\nu^{(3)}}J_\rho^{(2)}\right)~. \la{4.1}
\end{align}
Here $J_\pm = g^{-1} \partial_\pm g$ with $g = \diag(g_L,g_R) \in \grp{PSU}(1,1|2) \times \grp{PSU}(1,1|2)$ 
and $\STr$ is the supertrace, an invariant bilinear form on the Lie superalgebra $\alg{psu}(1,1|2) \oplus \alg{psu}(1,1|2)$. $P_{0,1,2,3}$ denote the projectors onto the $\Integer_4$ grading of $\alg{psu}(1,1|2)\oplus\alg{psu}(1,1|2)$, see, e.g., \cite{Seibold:2019dvf}, $P_F = P_1 + P_3$ is the projector onto the fermionic part of the superalgebra and $J^{(i)}_\pm = P_i J_\pm$. $W$ acts as $W\diag(X,Y) = \diag(X,-Y)$ where $\diag(X,Y) \in \alg{psu}(1,1|2)\oplus\alg{psu}(1,1|2)$, while the linear operators $d_-$ and $\Omega_-$ (that depends on the dressed R-matrix $R_g = \Ad_g^{-1} R \Ad_g$ where $\Ad_g$ denotes the standard adjoint action and $R$ is also assumed to satisfy $R^3 = -R$) are defined as
\begin{equation} \begin{aligned}\la{4.2}
d_- &= 2 P_2 + \frac{1}{1-k^2} \big( (\lambda-k^2) P_1 - (1+\lambda) k W P_3
-(\lambda+k^2) P_3 - (1-\lambda) k W P_1 \big)~,
\\
\Omega_- & = - \frac{\sqrt{(\mu-1)(1-k^2 \mu)}}{1+k W} R_g - k W \big( \frac{\mu-1}{1+k W}\big)\, R_g^2~,
\end{aligned}
\end{equation}
where
\begin{equation}\la{4.4}
\lambda = \sqrt{\frac{(1-k^2-\eta_L^2)(1-k^2-\eta_R^2)}{1-k^2}}~, \qquad \mu = 1+\frac{1}{\lambda^2+k^2} \diag(\eta_L^2,\eta_R^2)~.
\end{equation}
The string tension $T$ enters \rf{4.1} as an overall factor.
The deformation parameters $\eta_L$ and $\eta_R$ are expected to be associated with a quantum group deformation of the left and right copies of $\alg{psu}(1,1|2)$, with standard deformation parameters ${\rm q}_{L,R}$
(see, e.g., \ci{Hoare:2021dix} and references there). 
In the case $\eta_R = 0$, i.e. ${\rm q}_R = 1$, we give a conjecture for the deformation parameter ${\rm q}_{L}$ in terms of the remaining parameters in the action \rf{4.1} in Appendix \ref{app:dist}. 
The final parameter $k$ is related to the presence of the Wess-Zumino term in the second line of~\rf{4.1}, which although defined in 3d gives a local coupling in the 2d sigma model.
As usual, since $\grp{PSU}(1,1|2)$ has an $\grp{SU}(2)$ subgroup, the corresponding level ${\rm k}=T k$ is integer-quantized.
In the mixed flux model without the quantum group deformation (i.e.~$\eta_L=\eta_R=0$), the $k=0$ point corresponds to pure RR flux, while the $k=1$ point to pure NSNS flux.
The Lax connection demonstrating that the action \rf{4.1} defines a classically integrable sigma model is given in \cite{Delduc:2018xug}.

Denoting by $P_L$ and $P_R$ the projectors onto the left and right copies of $\alg{psu}(1,1|2)$, if we set $\eta_L=0$ (respectively $\eta_R=0$) then $P_L \Omega_- P_L=0$ (respectively $P_R \Omega_- P_R=0$).
This implies that the deformed action~\eqref{4.2} is invariant under the global left action $g \rightarrow g_0 g$ where $g_0 = \diag(g_{0L},1)$ (respectively $g_0 = (1,g_{0R})$).
Therefore, only the right (respectively left) copy of $\alg{psu}(1,1|2)$ is deformed, hence half of the original 16 supersymmetries are preserved.
\unskip\footnote{Recall that the DJ R-matrix only commutes with the action of the Cartan subgroup, hence will generically break or deform all of the associated supersymmetries.} 

This provides a natural candidate to connect with the backgrounds discussed in section \ref{sec:2}.
Without loss of generality, we focus on the case $\eta_R=0$ for which
$\lambda$ in
\rf{4.4} is $\sqrt{1-k^2-\eta_L^2}$.
We find it useful to define
\be \la{466}
\kappa =
\frac{ \eta_L}{\sqrt{1-\eta_L^2}}\ .
\ee
To ensure Weyl invariance we take $R$ to be a DJ R-matrix built from a Cartan-Weyl basis with all fermionic simple roots.
The associated supergravity background, which we call the DJ background,
can be extracted following the methods described in \cite{Seibold:2019dvf} (see also \cite{Borsato:2016ose}).
Choosing an appropriate parametrisation for $g$, the metric is given by \eqref{4}. 
The NSNS and RR fluxes are of the form \eqref{22} with coefficients\footnote{Note that to make this matching explicit we rescale $T \to (1+\kappa^2)T$ and the 4-torus coordinates $x_r \to \frac{x_r}{\sqrt{1+\kappa^2}}$.\label{trescale}
Recall that the NSNS and RR fluxes scale as $H_3 \sim T$, $F_3 \sim T$ and $F_5 \sim T^2$. After this rescaling the WZ level is ${\rm k} = (1+\ka^2)T k$. }${}^{,}$\footnote{To quadratic order in the fermions, the RR fluxes and the dilaton $\Phi$ only appear in the GS action through the combination $\mathcal F = e^{\Phi} F$.
To determine the dilaton we require that the supergravity equations of motion are satisfied.
That this is possible is ensured by the unimodularity of the R-matrix.
In this case, the dilaton indeed turns out to be constant.}
\begin{equation}
\label{eq:yspecial}
\begin{aligned}
\mathbf{z}_1 &= (\sqrt{1+\kappa^2} \sqrt{1-q^2}, - \sqrt{\frac{1+\kappa^2}{q^2+\kappa^2}} q^2, -\sqrt{\frac{1+\kappa^2}{q^2+\kappa^2}} q \kappa,0,0)~, \\
\mathbf{z}_2 &= (0, - \sqrt{\frac{1+\kappa^2}{q^2+\kappa^2}} \kappa^2, \sqrt{\frac{1+\kappa^2}{q^2+\kappa^2}} q \kappa,0,0)~, \qquad q = \sqrt{1 - k^2 (1 + \kappa^2) }~.
\end{aligned}
\end{equation}
Here $q$ is related to $k$ in \rf{4.1}
so the
parameters are $q\in [0,1]$ and $\kappa \in [0,\infty)$ with
${\rm k}= T \sqrt{1+\kappa^2}\sqrt{1-q^2}$ being the Wess-Zumino level. Notice that $\kappa=0$ gives the mixed flux background \eqref{3}, while $q=1$ gives the RR deformation of \cite{Seibold:2019dvf} with equal deformation parameters.\footnote{With rescaled tension and torus coordinates.}
For comparison, we provide in Appendix \ref{app:dist} the background associated to a non-unimodular R-matrix, which only satisfies a set of generalised supergravity equations of motion.

At this point we recall that only a subset of the background fluxes \rf{22}--\rf{24} can be obtained from the undeformed mixed flux model by TsT transformations \rf{34}.
This subset is guaranteed to be classically integrable.
To generate the full set of background fluxes we also need to allow S-dualities.
The key point is that the DJ background \rf{eq:yspecial}, which defines a classically integrable string sigma model by construction, is not of the form \rf{34}.
It follows that starting from the DJ background and applying TsT transformations it is possible to generate the full set of background fluxes, thereby demonstrating the classical integrability of the string sigma model.

The DJ background \eqref{eq:yspecial} has one free parameter, $q$, in addition to $\kappa$. Let us now show that it is possible to generate six additional free parameters using only TsT transformations. As a first step it is useful to perform an $\grp{SO}(4)$ rotation to bring \eqref{eq:yspecial} to
\begin{gather} \label{eq:yspecial2}
\textbf{z}_1 = (\sqrt{1+\ka^2}\sqrt{1-q^2},q\sqrt{1+\ka^2},0,0,0) ~,
\qquad
\textbf{z}_2 = (0,0,-\ka\sqrt{1+\ka^2},0,0) ~.
\end{gather}
This can equivalently be obtained through an appropriately chosen TsT in the torus directions $x^6$ and $x^7$. One can then do a rotation $\mathcal R \in \grp{SO}(3)/\grp{SO}(2)$ that leaves $\mathbf{z}_1$ invariant but introduces two new parameters in $\mathbf{z}_2$ (this is equivalent to rotating the torus coordinates). Then, one does another rotation $\mathcal R \in \grp{SO}(4)/\grp{SO}(3)$ introducing three new parameters in $\mathbf{z}_1$ (equivalent to TsT transformations). At this point it is useful to write the resulting vectors as
\begin{equation}
\mathbf{z}_1 = (\sqrt{1+\kappa^2} \sqrt{1-q^2}, \mathbf{u}_1)~, \qquad \mathbf{z}_2 = (0,\mathbf{u}_2)~,
\end{equation}
where the vectors $\mathbf{u}_1 = (y_1, y_3, y_5, y_7)$ and $\mathbf{u}_2 = (y_2, y_4, y_6, y_8)$ satisfy the constraints $\lVert \mathbf{u}_1 \rVert^2 = q^2 (1+\kappa^2)$, $\lVert \mathbf{u}_2 \rVert^2 = \kappa^2 (1+\kappa^2)$ and $\mathbf{u}_1 \cdot \mathbf{u}_2 = 0$. There are five independent parameters in addition to $\kappa$ and $q$. Notice that the first components of $\mathbf{z}_1$ and $\mathbf{z}_2$ are left invariant under the above rotations, which traduces the fact that we only do transformations in the torus directions, and no S-duality rotation. The last step consists of obtaining a non-trivial $s_2$ (the first component in $\mathbf{z}_2$). To achieve this we use a TsT in the left Cartan directions, as discussed in \ref{cart}. This leads to
\begin{equation}
\hat{\mathbf{z}}_1 = (\sqrt{1+\kappa^2} \sqrt{1-q^2}, \hat{\mathbf{u}}_1)~, \qquad \hat{\mathbf{z}}_2 = (\sqrt{\hat{\kappa}^2-\kappa^2} \sqrt{\hat{\kappa}^2+q^2} ,\hat{\mathbf{u}}_2)~,
\end{equation}
with the constraints $\lVert \hat{\mathbf{u}}_1 \rVert^2 = \hat{\kappa}^2 - \kappa^2 + q^2 (1+\kappa^2)$, $\lVert \hat{\mathbf{u}}_2 \rVert^2 = -q^2 (\hat{\kappa}^2 -\kappa^2) + \hat{\kappa}^2 (1+\kappa^2)$ and $\hat{\mathbf{u}}_1 \cdot \hat{\mathbf{u}}_2 = - \sqrt{1+\kappa^2} \sqrt{1-q^2} \sqrt{\hat{\kappa}^2-\kappa^2} \sqrt{\hat{\kappa}^2 + q^2}$. This is equivalent to the supergravity constraints \eqref{24}, except with $\hat{\kappa}$ on the right-hand side.

\section{Analytically-continued solution and special limits}
\la{sec:cont}

Thus far we have considered the deformed \adss solution with metric \rf{4} and fluxes \rf{22}--\rf{24}.
In particular, the metric on $\AdS_3$ is written in global coordinates and in \rf{8} as a time-like fibration over $\Hy^2$.
Our main motivation for this is that this is the metric that follows from the $\eta_L$ deformation of the $\Integer_4$ permutation supercoset with WZ term as discussed in section \ref{sec:int}, thereby allowing us to prove classical integrability.

Here we consider different transformations of the deformed background that still give deformations of \adss defining integrable string sigma models, albeit in different patches of $\AdS_3$. 
This includes the analytic continuation $\ka = i \td\ka$, which gives a deformation of $\AdS_3$ written as a space-like fibration over $\AdS_2$, as well as different scaling limits that give deformations in Poincar\'e patch.
We will also discuss the Schr\"odinger space-time
and pp-wave limits of the background in section \ref{sec:2}.

\subsection{Analytically-continued background}\label{acb}

Given the deformed \adss solution with metric \rf{4} and fluxes \rf{22}--\rf{24}
one may formally consider the case of $\ka^2 <0$ or set $\ka = i\td\ka$.
Then the metric \rf{4} remains real, but the final condition in \rf{24}
implies that at least one of the parameters in ${\bf z}_2$ in \rf{23} should become imaginary (we will assume that $\tilde{\kappa}^2 < 1$). 
Since ${\bf z}_2$ is the coefficient of $\extder \CC$ in \rf{20}, the only way to preserve the reality of fluxes
is to set ${\bf z}_2 =- i \td{\bf z}_2 $ and compensate this by an analytic continuation of the coordinates
so that $\CC$ becomes imaginary.
We will set
\be \la{36}
\kappa = i\td\kappa, \ \ \qquad \textbf{z}_2 = - i \td{\textbf{z}}_2 , \qquad
\psi = i \td t ~, \qquad t = i \td \psi ~, \ee
which leads to the following metric
\begin{equation} \la{37}
\begin{aligned}
\extder s^2 & = (1+\rho^2) \extder \tilde \psi^2 + \frac{\extder \rho^2}{1+\rho^2} - \rho^2 \extder\tilde t^2 - \tilde \kappa^2 \big((1+\rho^2) \extder \tilde\psi - \rho^2\extder \tilde t\big)^2 \\
&\quad+ (1-r^2) \extder \varphi^2 + \frac{\extder r^2}{1-r^2} + r^2 \extder \phi^2 - \tilde\kappa^2 \big((1-r^2) \extder \varphi + r^2\extder \phi\big)^2 + \extder x_r\extder x_r~.
\end{aligned}
\end{equation}
Note that, setting $\td \ka = 0$,
the analytically-continued metric still describes $\AdS_3$, however the coordinates $\td t, \td \psi$ and $\rho$
are no longer global coordinates.\foot{For $\td \k=0$ this metric follows from
$\extder s^2_{3} =-\extder X_{-1}^2 - \extder X_0^2 + \extder X_1^2 + \extder X_2^2 $,
with $X_{-1} \pm X_1 = \sqrt{1+\rho^2} e^{\pm \tilde{\psi}}$ and $X_2 \pm X_0 = \rho\, e^{\pm \tilde{t}}$.
Here $X_2^2 - X_0^2 > 0$ and $X_{-1} > 0$. The other inequalities $X_{-1}^2 - X_1^2 > 1$ and
$(X_{-1}^2 - X_1^2) - (X_2^2 - X_0^2) > 0$ follow from the first one and the embedding constraint
$-X_{-1}^2 - X_0^2 + X_1^2 + X_2^2 = -1$.}

The corresponding metric in Hopf parametrization found from \rf{7}, \rf{8} and \rf{36}
is ($\zeta_{1} = i \td\zeta_{1}$, $\zeta_2 = - i \td\zeta_2$)\foot{Closely related backgrounds were considered in \cite{Orlando:2010ay,Orlando:2012hu}.
There the deformed metric was generated by applying TsT transformations in $\alg{su}(1,1)_L \oplus \alg{u}(1)_{\To^4}$ and $\alg{su}(2)_L \oplus \alg{u}(1)_{\To^4}$ where $\alg{u}(1)_{\To^4}$ denotes shift isometries associated to the 4-torus.
This gave the metric \eqref{37}, also allowing for different deformation parameters in the $\AdS_3$ and $\Sp^3$ parts.
However, in contrast to the case we are considering here, the NSNS and RR 3-forms there
involved the torus directions.
This means that in the reduction to 6d (considered in section \ref{sols}) there are also non-zero
2-form field strengths.
Additionally, in the construction of \cite{Orlando:2010ay,Orlando:2012hu}, if $\AdS_3$ and $\Sp^3$
were both deformed no supersymmetries were preserved,
again in contrast to the background discussed here.}
\begin{equation} \la{40}
\begin{aligned}
{\extder s}^2 & = \tfrac{1}{4} \big( - \sinh^2 \bs \, \extder \td{\zeta}_2^2 + \extder \bs^2 + (1-\td \kappa^2) (\extder \td{\zeta}_1 + \cosh \bs\, \extder \td{ \zeta}_2)^2 \big) \\
&\quad + \tfrac{1}{4} \big( \sin^2 {\s} \, \extder {{\xi}}_2^2 + \extder {\s}^2 + (1-\td \kappa^2) (\extder {{\xi}}_1 - \cos {\s}\, \extder { {\xi}}_2)^2 \big)+ \extder x_r\extder x_r~,
\end{aligned}
\end{equation}
The fluxes supporting this metric are given by
\begin{equation} \la{338}
\begin{aligned}
H_3 &= s_1 \extder\hat{B}+ \tilde s_2 \extder \CCC ~, \qquad
F_3 = y_1 \extder\hat{B}+ \tilde y_2 \extder\CCC ~, \\
F_5 &= (y_3 \extder\hat{B} + \tilde y_4 \extder\CCC ) \wedge J_2^{(1)} + (y_5 \extder\hat{B} + \tilde y_6 \extder\CCC ) \wedge J_2^{(2)}+(y_7 \extder\hat{B} + \tilde y_8 \extder\CCC ) \wedge J_2^{(3)} ~,\\
\extder \hat{B} &= \tfrac{1}{4} \big[ \sinh \bs \, \extder \td\zeta_1 \wedge \extder \td\zeta_2 \wedge \extder \bs + \sin \s \, \extder \xi_1 \wedge \extder \xi_2 \wedge \extder \s\, \big]~,\\
\extder {\CCC} &= \tfrac{1}{4} \big[ - \sinh \bs\, \extder \td\zeta_2 \wedge \extder \bs\, \wedge (\extder \xi_1 - \cos \s\, \extder \xi_2) + \sin \s (\extder \td\zeta_1 + \cosh \bs\, \extder \td\zeta_2) \wedge \extder \xi_2 \wedge \extder \s \big]~,
\end{aligned}
\end{equation}
where $\textbf{z}_2 \extder \CC = \tilde{\textbf{z}}_2 \extder\CCC$.
The supergravity equations then imply that
$\textbf{z}_1 = (s_1,y_1,y_3,y_5,y_7)$ and $ \tilde{\textbf{z}}_2=(\tilde s_2, \tilde y_2, \tilde y_4, \tilde y_6, \tilde y_8)$ should satisfy (cf. \rf{24})
\begin{align}
\textbf{z}_1 \cdot \tilde { \textbf{z}}_2 = 0~, \qquad \lVert\textbf{z}_1\rVert^2= 1-\tilde \kappa^2 ~, \qquad
\lVert\tilde{\textbf{z}}_2\rVert^2= \tilde \kappa^2 (1-\tilde\kappa^2) ~ . \la{39}
\end{align}

While the AdS part of \rf{8} is a time-like fibration over $\Hy^2$, in the analytically-continued metric \rf{40} it is a space-like fibration over
AdS$_2$ with Lorentzian signature. The metric \rf{40} thus interpolates between
$\rm AdS_3 \times S^3 \times \To^4$ ($\tilde \kappa = 0$) and $\rm AdS_2\times S^2 \times \To^6$ ($\tilde \kappa = 1$).
In order to preserve the non-degeneracy of the metric \rf{40} in the limit $\td \k=1$
we also need to rescale the coordinates
\begin{equation}\la{41}
\td {\zeta}_1 \rightarrow \epsilon^{-1} \td {\zeta}_1~, \qquad \xi_1 \rightarrow \epsilon^{-1} \xi_1~,\qquad
\epsilon = \sqrt{1-\td \kappa^2}\ .
\end{equation}
It then follows from \rf{338} that $\extder \hat{B} \sim \extder \CCC \sim \epsilon^{-1}$.
To get a consistent solution with non-singular fluxes from \rf{40}--\rf{39} we need to simultaneously
rescale the parameters $\bz_1\to \bz_1'=\epsilon^{-1} \bz_1, \ \td \bz_2\to \bz_2'=\epsilon^{-1} \td \bz_1$
such that $\bz_1'$ and $\bz_2'$ become unit-normalised and orthogonal.

Explicitly, introducing the vielbein
\begin{equation}\begin{gathered} \la{43}
e'_1 =\ha \dd \tilde{\zeta}_1 ~,
\quad
e'_0=\ha \sinh \bs\, \extder \tilde{\zeta}_2 ~,
\quad
e_2= \tfrac{1}{2} \extder \bs ~,
\\
e_3 = \ha \dd {{ \x}}_1 ~, \quad
e_4= \ha \sinh {\s}\, \extder { \xi}_2 ~, \quad
e_5=\tfrac{1}{2} \extder {\s} ~,
\end{gathered}\end{equation}
we may write the resulting $\rm AdS_2 \times S^2 \times T^6$ metric and fluxes as
\begin{equation}
\begin{aligned}\la{44}
\extder s^2 &= ( -e'^2_0 + e_2^2 ) + (e_4^2 + e^2_5 ) + (e'^2_1 + e_3^2 + \dd x^r \dd x^r) \ , \\
H_3 &= s'_1 \extder \hat{B}' + s'_2 \extder {\CC'}~, \qquad
F_3 = y'_1 \extder \hat{B}' + y'_2 \extder {\CC'}~, \\
F_5 &= ( y'_3 \extder \hat{B}' + y'_4 \extder {\CC'}) \wedge J_2^{(1)}+( y'_5 \extder {\hat B}' + y'_6 \extder {\CC'}) \wedge J_2^{(2)}+( y'_7 \extder {\hat B}' + y'_8 \extder {\CC'}) \wedge J_2^{(3)} ~, \\
\extder \hat{B}' &= 2 \left( e'_0 \wedge e'_1 \wedge e_2 + e_3 \wedge e_4 \wedge e_5 \right)~, \qquad
\extder {\CC}' = 2 \left( - e'_0 \wedge e_2 \wedge e_3 + e'_1 \wedge e_4 \wedge e_5 \right)~,
\end{aligned}
\end{equation}
where $ \textbf{z}_1' = (s'_1,y'_1,y'_3,y'_5,y'_7)$ and $ \textbf{z}_2'=(s'_2,y'_2,y'_4,y'_6,y'_8)$ satisfy
\be
\textbf{z}'_1 \cdot { \textbf{z}}'_2 = 0~, \qquad \lVert\textbf{z}'_1\rVert^2= 1 ~, \qquad
\lVert{\textbf{z}}'_2\rVert^2= 1 ~ . \la{42}\ee
If we choose ${s}_1' = {s}_2' = {y}_1' = {y}_2' =0$ such that only the $F_5$ flux is non-zero
then the remaining parameters should satisfy
\begin{equation} \la{49}
y'_3 y'_4+y'_5 y'_6+y'_7 y'_8 =0~, \qquad
y'^2_3 + y'^2_5 + y'^2_7 =1~, \qquad
y'^2_4 + y'^2_6 + y'^2_8 =1~.
\end{equation}
This gives a three-parameter family of $\rm AdS_2 \times S^2 \times T^6$
type IIB solutions supported by
\begin{equation}\la{50}
F_5 \sim \text{Vol}(\AdS_2 ) \wedge \Re \Omega_3 + \text{Vol}(\Sp^2 )\wedge \Im \Omega_3 ~,
\end{equation}
where $\text{Vol}( \AdS_2) = e_0\wedge e_2$, $\text{Vol}(\Sp^2 ) = e_4 \wedge e_5$ and
$\Omega_3 = \extder w^1 \wedge \extder w^2 \wedge \extder w^3$ is a holomorphic three-form on the 6-torus.\foot{This corresponds to a near-horizon limit of the type IIB solution representing a $\tfrac14$-supersymmetric
intersection of four D3-branes \ci{Klebanov:1996mh,Sorokin:2011rr}.}

Let us note that
starting with the background \rf{40}--\rf{39} for general values of the parameters and compactifying to 4 dimensions along the isometric directions
$ \td{\zeta}_1$ and $\xi_1$ as well as $\To^4$ gives a family of \adst solutions of $d=4$ supergravity only supported
by several equal-charge electric and magnetic Maxwell fluxes.
One pair of 4d abelian vector fields come from the KK reduction on the fibres
and others come from the five 3-forms in the 6d action \rf{15}. The effective 4d Lagrangian is then
$L= R_4 - \tfrac14 \sum_k c_{k} F^{(k)}_{uv} F^{(k)uv} $ (cf. \ci{Lu:1997hb,Duff:1998cr,Cvetic:2021lss}).
These \adst solutions may be viewed as the near-horizon limits of a family of $d=4$ $\N=2$ supersymmetric
extremal RN black holes (with constant scalar fields).\foot{In general,
one can use U-duality to represent the BPS
4d black hole solution (with non-constant scalars) as having 5 charges all in the NSNS sector \ci{Cvetic:1995bj,Chan:1996ia}
but such solutions need not uplift just to our effective background \rf{44} or its 6d reduction corresponding to \rf{15}.
The reason is that we have not considered TsT transformations that change the structure of the 6d metric,
i.e.~there is effectively more freedom from the 4d perspective.
Thus these extra T-dualities may allow one to put the metric into the form
in which the most general solution can be generated from the one with NSNS charges only.}

As discussed in section \ref{sec:int},
the background \rf{8} and \rf{22} corresponds to an integrable GS sigma model,
and thus the same applies also to its analytic continuation \rf{37} and \rf{338}.
The above relation
may therefore be interpreted as an integrable embedding of these \adst backgrounds into type IIB string theory.

\subsection{Scaling backgrounds}\label{sca}

Another transformation of interest is the scaling limit
\begin{equation}\label{eq:scal}
\sigma \to \log \frac{2z}{\epsilon} ~, \qquad \zeta_2 \to \epsilon \zeta_2 ~, \qquad \epsilon \to 0 \ .
\end{equation}
Taking this limit in the metric \rf{8} and fluxes \rf{99} we find
\begin{equation} \label{scal1}
\begin{aligned}
\extder s^2 &= \tfrac{1}{4} \big( z^2 \extder \zeta_2^2 + z^{-2}\extder z^2 - (1+\kappa^2) (\extder \zeta_1 - z\,\extder \zeta_2)^2 \big)
\\
&\qquad +\tfrac{1}{4} \big( \sin^2 \s \, \extder \xi_2^2 + \extder \s^2 + (1+\kappa^2) (\extder \xi_1 - \cos \s\, \extder \xi_2)^2 \big)~.
\\
\extder \hat{B} &= \tfrac{1}{4} \big[ \extder \zeta_1 \wedge \extder \zeta_2 \wedge \extder z + \sin \s \, \extder \xi_1 \wedge \extder \xi_2 \wedge \extder \s\, \big]~,\\
\extder {\CC} &= \tfrac{1}{4} \big[ \extder \zeta_2 \wedge \extder z\, \wedge (\extder \xi_1 - \cos \s\, \extder \xi_2) + \sin \s (\extder \zeta_1 - z \, \extder \zeta_2) \wedge \extder \xi_2 \wedge \extder \s \big]~.
\end{aligned}
\end{equation}
Since we are not taking any limit on $\ka$ this metric can be completed to supergravity solutions by the ans\"atze \rf{22} subject to the constraints \rf{24}.

Setting $\ka = 0$ in the AdS part of the metric~\eqref{scal1} we find
\begin{equation}\label{tlpp}
\extder s^2_{_{\rm AdS}} = \tfrac14 \big(z^{-2}\extder z^2 + 2z \extder \zeta_1 \extder \zeta_2 - \extder \zeta_1^2) ~,
\end{equation}
which we recognise as the background of a
pp-wave in $\AdS_3$ and is locally equivalent to $\AdS_3$ \cite{Brecher:2000pa,Kruczenski:2008bs}.\foot{Starting from the metric of $\AdS_3$ in Poincar\'e patch $\extder s^2 = \td z^{-2} (\extder \td z^2 + 2 \extder u \extder v)$ and using the change of coordinates $\td z = z^{-\frac12}\sec\frac{\mu\zeta_1}{2}$, $u = \mu^{-1}\tan\frac{\mu\zeta_1}{2}$, $v=\frac12(\zeta_2 - \mu z^{-1}\tan\frac{\mu\zeta_1}{2})$, we find the metric~\eqref{tlpp}.
For $\mu = i$, which still defines a real coordinate transformation, we find the space-like pp-wave in $\AdS_3$, i.e.~\eqref{tlpp} with $\zeta_1 = i \tilde\zeta_1$, $\zeta_2 = - i \tilde\zeta_2$.}
Furthermore, as well as being related by the scaling limit~\eqref{eq:scal},
the metric \rf{8} and fluxes \rf{99} for general $\ka$ are also related to \eqref{scal1} by just the following
local coordinate redefinition
\begin{equation}\begin{gathered}\label{cf}
\sinh\frac{\sigma}{2} \to \sqrt{\frac{1-4z +z^2 (4+\zeta_2^2)}{8z}} ~, \qquad
\sin2\zeta_2 \to -\frac{8z^2 \zeta_2(1-z^2(4-\zeta_2^2))}{1-2z^2(4-\zeta_2^2)+z^4(4+\zeta_2^2)^2} ~,
\\
\sin2\zeta_1 \to \frac{(1-2z^2(4+3\zeta_2^2)+z^4(4+\zeta_2^2)^2)\sin 2\zeta_1 - 4 z \zeta_2(1-z^2(4+\zeta_2^2))\cos 2\zeta_1}{(1-4z+z^2(4+\zeta_2^2))(1+4z+z^2(4+\zeta_2^2)}~.
\end{gathered}
\end{equation}
This means that the scaling limit does not change the form of the YB deformed supercoset action \rf{4.1} or the unimodular R-matrix -- it only modifies the particular parametrisation of the supergroup-valued field $g$ that is used to find the form of the background in local coordinates.\footnote{Explicitly, considering just the $\AdS_3$ sigma model, or equivalently the $\grp{SU}(1,1)$ PCM, if we parametrise $g\in \grp{SU}(1,1)$ as
$g = \exp\big(-\tfrac{i\zeta_2}{2}\sigma_3\big) \exp\big(\frac{\sigma}{2}\sigma_2\big) \exp\big(\frac{i\zeta_1}{2}\sigma_3\big)$ with $R(X) = \tfrac12 (\tr(\sigma_2 X)\sigma_1 - \tr(\sigma_1 X)\sigma_2)$ we recover the $\AdS$ part of the metric \rf{8} from the YB deformation of the PCM.
Alternatively, if we parametrise
$g = \tfrac{1}{2\sqrt{2z}} \big((1+2z) 1_2 - z \zeta_2 \sigma_1 - (1-2z) \sigma_2 - i z \zeta_2\sigma_3\big)\exp\big(\frac{i\zeta_1}{2}\sigma_3\big)$, then using the same R-matrix we recover the $\AdS$ part of the metric \rf{scal1}.}

We can also take the analogous limit to \eqref{eq:scal}, i.e.~with $\tilde \zeta_2 \to \epsilon \tilde \zeta_2$, in the analytically-continued metric \rf{40} and fluxes \rf{338}.
The resulting metric and fluxes are the same as \eqref{scal1} with $\zeta_1 = i \tilde\zeta_1$, $\zeta_2 = - i \tilde\zeta_2$, $\kappa = i \tilde \kappa$ and $\CC = i \CCC$, and this metric can be completed to supergravity solutions by the ans\"atze \rf{338} subject to the constraints \eqref{39}.
Setting $\td\ka = 0$ we find the metric
$
\extder s^2_{_{\rm AdS}} = \tfrac14 \big(z^{-2}\extder z^2 + 2z \extder \td \zeta_1 \extder\td \zeta_2 + \extder \td \zeta_1^2) $
(which is a limit of the 3d F1+ppwave background \ci{Horowitz:1994rf}).\foot{This is, in fact, how the AdS$_3$ metric appears in the near-horizon limit of the F1+NS5+pp-wave+KK-monopole
sigma model providing a particular string embedding of 4d BPS black holes in \ci{Cvetic:1995yq,Cvetic:1995bj}
(see eq. (17) in \ci{Cvetic:1995yq} and eq. (20) in \ci{Cvetic:1995bj}).}
Again this metric is locally equivalent to $\AdS_3$ and, moreover, the deformed metric and fluxes are related to \eqref{40} and \rf{338} by a local coordinate redefinition similar to \rf{cf}.
Similar deformations to this analytically-continued pp-wave background and their holographic interpretation have been studied, e.g., in \cite{El-Showk:2011euy,Bena:2012wc}.

\subsection{Special limit leading to Schr\"odinger background}\la{jord}

There is a particular limit
of the deformed background \rf{8}, \rf{22} that
gives rise to the metric which is a direct sum of that of the 3d Schr\"odinger space-time and undeformed $\rm S^3$ (and 4-torus).
Indeed, let us rescale
\begin{equation} \la{517}
\kappa \to \epsilon \kappa~, \quad
\textbf{z}_2 \to \epsilon \textbf{z}_2 ~, \qquad (\zeta_1,\zeta_2) \to \epsilon(\zeta_1,\zeta_2)~, \quad \sigma \to \sigma - \log \epsilon^{2}~,
\end{equation}
and take $\epsilon \to 0$.
Then the metric in \rf{8} and the auxiliary fluxes in \rf{99}, \rf{20} become\foot{The same background can be
found by taking an analogous limit, i.e.~with ${\bf z}_2 \to - \epsilon {\bf z}_2$ and $(\tilde\zeta_1,\tilde\zeta_2) \to \epsilon(\tilde\zeta_1,\tilde\zeta_2)$, in the analytically-continued background \rf{40}--\rf{338}.
It can also be found from the scaled background~\rf{scal1} by setting $\ka \to \epsilon\ka$, ${\bf z}_2 \to \epsilon {\bf z}_2$, $\z_1 \to\epsilon \zeta_1$, $\zeta_2 \to \epsilon^{-1} \zeta_2$ and $z \to \tfrac12 e^{\sigma}$ and taking $\epsilon \to 0$, or
from
its analytic continuation by an analogous limit (cf. \ci{El-Showk:2011euy,Bena:2012wc}).}
\begin{align} \no
&\extder s^2 = \tfrac{1}{4} \big( \extder \sigma^{2} + e^{\sigma} \extder \zeta_1 \extder \zeta_2 -\tfrac{1}{4} \k^2e^{2\sigma} \extder \zeta_2^2 \big)
+\tfrac{1}{4} \big( \sin^2 \s \, \extder \xi_2^2 + \extder \s^2 + \left(\extder \xi_1 - \cos \s\, \extder \xi_2\right)^2 \big) + \extder x_r \extder x_r ~,
\\\label{eq:jord}
&\bF_3=\textbf{z}_1 \extder \hat{B} + \textbf{z}_2 \extder {\CC} = \textbf{z}_1 \big(\tfrac18 e^{\sigma} \extder \zeta_1 \wedge \extder \zeta_2 \wedge \extder \sigma
+ \tfrac14\sin \theta \extder \xi_1 \wedge \extder \xi_2 \wedge \extder \theta\big)
\\ \no
& \qquad \qquad \qquad\qquad \ \ \ \ + \textbf{z}_2 \big(\tfrac18 e^{\sigma} \extder \zeta_2 \wedge \extder \sigma \wedge (\extder \xi_1 - \cos\theta \extder \xi_2) - \tfrac18 e^{\sigma} \sin\theta \, \extder \zeta_2 \wedge \extder \xi_2 \wedge \extder \theta\big) ~.
\end{align}
Here the first 3d factor of the metric is that of the
3d Schr\"odinger space-time (equivalent to $ \tfrac{\extder\rz^2}{\rz^2} + \frac{1}{4\rz^2} \extder \zeta_1 \extder \zeta_2 -
\frac{\k^2}{16\rz^4} \extder \zeta_2^2$ where $\rz= e^{-\sigma/2}$) which is
a deformation of $\AdS_3$ directly in the Poincar\'e patch.

The corresponding limit of the supergravity constraints \rf{24} is
\begin{equation}\la{519}
\textbf{z}_1 \cdot \textbf{z}_2 =0~, \qquad \lVert\textbf{z}_1\rVert^2=1~, \qquad \lVert\textbf{z}_2\rVert^2=\kappa^2~.
\end{equation}
Let us note that it is possible to set $\ka = 1$ by rescaling $\zeta_1 \to \ka \zeta_1$, $\zeta_2 \to \ka^{-1} \zeta_2$ and $\textbf{z}_2 \to \ka \textbf{z}_2$, meaning that $\ka$ is not a genuine deformation parameter.

For the particular solution of \rf{519}
\begin{equation}\label{eq:jordsol}
\textbf{z}_1 = (\sqrt{1-q^2}, q,0,0,0) ~, \qquad
\textbf{z}_2 = (0,0,-\kappa,0,0) ~,
\end{equation}
this limit is related to a special
Jordanian deformation
limit
of the DJ background (with parameters given by \eqref{eq:yspecial2}).
Taking the limit in the YB deformed supercoset action \rf{4.1} is equivalent to
using, instead of the
DJ R-matrix,
a fermionic extension of the Jordanian R-matrix in \cite{vanTongeren:2019dlq,Hoare:2016hwh}.
Introducing the usual Cartan-Weyl basis of generators $H_L$, $E_L$ and $F_L$ for $\alg{sl}(2;\Real)_L$ and similarly for $\alg{sl}(2;\Real)_R$, the bosonic part of the R-matrix acts as $R(F_L) = - H_L$, $R(H_L) = 2 E_L$, and $R(E_L) = 0$.
This R-matrix solves the cYBe and as such corresponds to a homogeneous YB deformation \cite{Kawaguchi:2014fca} of the $\AdS_3 \times \Sp^3$ background. This implies
that the deformed background \rf{eq:jord}
can also be found from \adss by a non-abelian duality in a particular subalgebra of $\alg{psu}(1,1|2)$ \cite{Hoare:2016wsk,Borsato:2017qsx}.

Another case of the background~\eqref{eq:jord} can also be found by starting from the mixed flux $\AdS_3 \times \Sp^3$ background \rf{1}, \rf{3} in Poincar\'e coordinates, i.e.~\eqref{eq:jord} with \be
\ka = 0, \qquad \textbf{z}_1 = (\sqrt{1-q^2},q,0,0,0), \qquad
\textbf{z}_2 = \textbf{0} , \la{522}\ee
and applying a special TsT transformation in null coordinate $\zeta_1$ or null Melvin twist \cite{Matsumoto:2015uja}:
T-duality in $\xi_1$, shift $\zeta_1 \to \zeta_1 + \frac{4\ka}{q} \xi_1$ and T-duality back in $\xi_1$.
This leads to the background \rf{eq:jord} corresponding to the following solution to the constraints~\eqref{519}
\begin{equation}\label{eq:tstsol}
\textbf{z}_1 = (\sqrt{1-q^2},q,0,0,0) ~, \qquad
\textbf{z}_2 = (\ka q,-\ka \sqrt{1-q^2},0,0,0) ~.
\end{equation}

Just as for the deformation of $\AdS_3 \times \Sp^3$ in
global coordinates discussed in section \ref{sec:2}, the full space of solutions of~\eqref{519} can be found by starting from the Jordanian solution~\eqref{eq:jordsol} and using the above null Melvin twist and TsT transformations on the 4-torus.
This demonstrates the classical integrability of the corresponding string sigma model.
Equivalently, while the full space of solutions of \rf{519}
cannot be generated from \adss by just TsT transformations, it can
if we also allow a non-abelian duality transformation.
This is in contrast to the situation before taking the limit \rf{517}, where, as discussed in section \ref{sols},
to get the most general background \rf{8}, \rf{20} using duality transformations we also need to apply S-duality (see \rf{26}).\foot{Note that after taking limit we can still generate the full space of solutions of \rf{519} using TsT transformations and S-duality, i.e., without non-abelian duality.}

While the full space of solutions \rf{519} can be generated using non-abelian duality and TsT transformations,
this cannot be lifted to the F1-D1-NS5-D5 brane background \cite{Tseytlin:1997cs}
which, in the near-horizon limit, becomes the mixed flux $\AdS_3 \times \Sp^3$
background \rf{1}, \rf{3}.\foot{Explicitly, $J = H_\ind{R} - H_\ind{L}$, $P_+= E_\ind{R}$ and $P_- = E_\ind{L}$ are the generators of the Lorentz algebra, and
$D = H_\ind{R} + H_\ind{L}$, $K_- = F_\ind{R}$ and $K_+ = F_\ind{L}$ generate dilatations and special conformal transformations respectively.}
On the other hand, we can apply the null Melvin twist, TsT transformations on the 4-torus and S-duality to the brane background.
This generates a 8-parameter (including $\ka$) deformation, which in a modified near-horizon limit, where $\ka$ is also rescaled, gives the deformed $\AdS_3 \times \Sp^3$ background~\eqref{eq:jord}.

\subsection{Plane-wave limit and light-cone S-matrix}
\la{sec:pp}

The plane-wave analog of the $\k$-deformed \adss metric \eqref{4} is obtained by
redefining the coordinates and taking the Penrose-type limit $L\to \infty$
\begin{align}
& t = \mu x^+ + \frac{x^-}{\mu L^2}~, \qquad \varphi = \mu x^+ - \frac{x^-}{\mu L^2}~,\qquad
\psi = \psi'-\kappa^2 \mu x^+~, \qquad \phi = \phi'-\kappa^2 \mu x^+~, \no\\
&
\qquad\qquad
\rho = \frac{\sqrt{1+\kappa^2} }{L}\rho' ~, \qquad r = \frac{\sqrt{1+\kappa^2} }{L}r'~,\qquad
L \to \infty\, . \la{526}
\end{align}
We also should rescale the 4-torus coordinates $x_r = \frac{\sqrt{1+\kappa^2} }{L}x'_r$
and the overall factor of string tension $ T = \frac{L^2}{1+\kappa^2}T'$ (cf. \ci{Sfetsos:1994fc}).\footnote{To recall, if we restore the dependence on string tension $T$ then
the metric and fluxes scale with $T$ as follows: $
\extder s^2 \sim T, \, H_3 \sim T, \, F_3 \sim T,\, F_5 \sim T^2$. We shall set $T'=1$ after taking the limit.}
Here $\mu$ is an effective curvature scale parameter.
As a result, we get the following pp-wave metric~
\begin{equation}\la{529}
\begin{aligned}
\extder s^2 = &
-4 \extder x^+ \extder x^- - (1 + \k^2)^2 {\mu}^2 (\extder x^+)^2 (\rho'^2 + r'^2) \\
&\ \ \ + \extder \rho'^2 + \rho'^2 \extder \psi'^2 + \extder r'^2 + r'^2 \extder \phi'^2 + \extder x'_r \extder x'_r ~.
\end{aligned}
\end{equation}
Note that the deformation parameter $\k$ enters only through
$\hat{\mu}= (1+\kappa^2)\mu$, which (for $\hat \mu\neq0$) can be rescaled away by
$x^+ \to \hat \mu^{-1} x^+, \ x^- \to \hat \mu x^-$.
Thus the metric \rf{529} is equivalent to the pp-wave metric \ci{Blau:2002dy} found from the
undeformed $\AdS_3 \times \Sp^3 \times \To^4$.

In the limit \rf{526} the auxiliary potentials in \rf{2}, \rf{6} differ only by an exact 2-form
\begin{align}\la{530}
& \hat{B}' = \mu\, \extder x^+ \wedge (\rho'^2 \extder \psi' + r'^2 \extder \phi')~, \qquad
{\CC}' = -\tfrac{2}{1+\kappa^2} \extder x^+ \wedge \extder x^- +\mu\, \extder x^+ \wedge (\rho'^2 \extder \psi' + r'^2 \extder \phi')\ ,\no \\
&\quad \qquad \extder \hat{B}' = \extder \CC' =2 \mu\, \extder x^+ \wedge (\rho' \extder \rho' \wedge \extder \psi' + r' \extder r' \wedge \extder \phi')~,
\end{align}
Note that in
terms of the four cartesian coordinates $z_i$ defined as
$z_1 + i z_2 = \rho\, e^{i \psi}$ and $z_3 + i z_4 = r\, e^{i \phi}$, the metric~\eqref{529} and 3-form~\rf{530} may be written as
\be \la{mm}
\extder s^2 = -4 \extder x^+ \extder x^- - (1+ \k^2)^2 {\mu}^2 (\extder x^+)^2 z_i^2 + \extder z_i^2 + \extder x'^2_r, \qquad \extder \hat{B}' = -2 \mu \extder x^+ \wedge (\extder z_1 \wedge \extder z_2 + \extder z_3 \wedge \extder z_4)\ . \ee
The flux background \eqref{22} supporting the metric \rf{529} thus becomes
\begin{equation} \la{531}
\begin{aligned}
& H_3 = (s_1+s_2) \extder \hat{B}'~, \qquad
F_3 = (y_1+y_2) \extder \hat{B}'~, \\
F_5 &= \extder \hat{B}' \wedge \big( (y_3+y_4) J'^{(1)}_2+(y_5+y_6) J'^{(2)}_2+(y_7+y_8)J'^{(3)}_2 \big)~,
\end{aligned}
\end{equation}
i.e.~${\bF}_3 = (\textbf{z}_1 + \textbf{z}_2) \extder \hat{B}'$, where the only condition on the parameters is
(cf. \rf{20}, \rf{24})
\be\la{532}
\lVert \textbf{z}_1 + \textbf{z}_2\rVert^2= (s_1+s_2)^2 + (y_1+y_2)^2 + (y_3+y_4)^2 + (y_5+y_6)^2 + (y_7+y_8)^2=
(1+\kappa^2)^2 \ .
\ee
Thus the background in this limit only depends on the parameters ${\bf z}_1$ and ${\bf z}_2$ through ${\bf z}_1 + {\bf z}_2$.
Therefore, two different instances of the original background~\rf{20} will have the same pp-wave limit if they have equal ${\bf z}_1 + {\bf z}_2$, but will differ at $O(L^{-2})$.

Starting with the superstring action corresponding to the deformed background \rf{4}, \rf{22} and considering the BMN-type
expansion
\ci{Berenstein:2002jq} in a light-cone gauge as in the undeformed case \ci{Hoare:2013pma,Hoare:2013lja} at the leading (quadratic) order we find the string propagating in the pp-wave background \rf{529}, \rf{531}.
The resulting 2d dispersion relation $ \omega(p)$ for the bosonic string fluctuations originating from the $\AdS_3 \times \Sp^3$ part of the model is\foot{As expected, setting $\ka = 0$ this dispersion relation matches the small-momentum limit of the conjectured exact dispersion relation for strings on $\AdS_3 \times \Sp^3$ with mixed flux \cite{Hoare:2013lja,Lloyd:2014bsa}. For $s_1=q (1+\kappa^2)$ and $s_2=0$ it matches the dispersion relation of the three parameter deformation \cite{Bocconcello:2020qkt} with equal quantum group deformation parameters. }
\begin{equation}\la{533}
(\omega- \varsigma \kappa^2)^2 = p^2 + 2 \varsigma (s_1+s_2)p + (1+\kappa^2)^2~, \qquad \varsigma \in \{-1,+1\}~.
\end{equation}
This dispersion relation becomes linear
for the special choice of $s_1+s_2$
\be \la{534} s_1+s_2 = \pm (1+\kappa^2) \quad \Rightarrow \quad
| \omega- \varsigma \kappa^2 |= | p \pm \varsigma (1+\kappa^2)| \ . \ee
It follows from \rf{532} that in this case the pp-wave background is pure NSNS, i.e.~$y_k+y_{k+1}=0$ for $k$ odd, which implies $F_3=F_5=0$.
However, if we set $s_1+s_2 = \pm (1+\kappa^2)$ and $y_k+y_{k+1}=0$ in the original
background \rf{20} (before taking the limit \rf{526}) then the conditions \eqref{24} imply that $s_1 = \pm 1$, $s_2 = \pm \kappa^2$ and $\sum_k y_k^2 = \kappa^2$. Hence, it is only pure NSNS, i.e.~$y_k=0$ for all $k$, if $\k=0$.
This corresponds to undeformed $\AdS_3 \times \Sp^3 \times \To^4$ supported by $H_3$ only.
While all backgrounds with $s_1+s_2 = \pm (1+\kappa^2)$ and $y_k+y_{k+1}=0$ have the same pure NSNS pp-wave limit, they will differ at the next (quartic) order in the BMN-type expansion.

Fixing a light-cone gauge in the $x^+$ direction (the expansion is around the $t=\varphi \sim \tau$ massive geodesic, cf. \rf{526}) in the full GS action we can compute the corresponding 4-point tree-level
S-matrix for the bosonic
fluctuations. It is convenient to introduce the complex scalar fields
\begin{equation}\la{5311}
Z = \frac{z_2 - i z_1}{\sqrt{2}}~, \qquad \bar{Z} = \frac{z_2 + i z_1}{\sqrt{2}}~, \qquad Y = \frac{z_3 - i z_4}{\sqrt{2}}~, \qquad \bar{Y} = \frac{z_3+i z_4}{\sqrt{2}}~,
\end{equation}
for the 2+2 transverse coordinates on $\AdS_3$ ($Z$ and $\bar{Z}$) and $\Sp^3$ ($Y$ and $\bar{Y}$) respectively
(cf. \rf{mm}). Assuming that $s_1+s_2 = 1+\kappa^2$ and $\left.\frac{\partial \omega}{\partial p} \right|_{p_1} >0$ and $\left.\frac{\partial \omega}{\partial p} \right|_{p_2} <0$,
we find
the following generalization of the pure NSNS \adss S-matrix (cf.~\cite{Hoare:2013pma,Hoare:2013lja,Lloyd:2014bsa,Baggio:2018gct})
to the case of $\kappa\neq 0$:\foot{
$S$ here stands for the tree-level T-matrix and we omit an overall factor proportional to the effective coupling (inverse string tension).}
\begin{itemize}
\item Left-Left sector ($\varsigma_1=\varsigma_2=-1$)
\begin{align}\no
S_{ZZZZ} &= -S_{YYYY} = -\tfrac{1}{2}(p_1+p_2) \frac{1+\kappa^2-p_2 (s_1-s_2+\kappa^2)}{1+\kappa^2-p_2} +\kappa^2 \frac{1+\kappa^2-p_1}{1+\kappa^2-p_2} p_2 ~, \\
S_{ZYZY} &= -S_{YZYZ} = \tfrac{1}{2}(p_1-p_2) \frac{1+\kappa^2-p_2 (s_1-s_2+\kappa^2)}{1+\kappa^2-p_2} - \kappa^2\frac{1+\kappa^2-p_1}{1+\kappa^2-p_2} p_2 ~,\la{535}
\end{align}
\item Right-Right sector ($\varsigma_1=\varsigma_2=+1$)
\begin{align}\no
S_{\bar{Z}\bar{Z}\bar{Z}\bar{Z}} &= -S_{\bar{Y}\bar{Y}\bar{Y}\bar{Y}} = \tfrac{1}{2}(p_1+p_2) \frac{1+\kappa^2+p_2 (s_1-s_2+\kappa^2)}{1+\kappa^2+p_2} - \kappa^2\frac{1+\kappa^2+p_1}{1+\kappa^2+p_2} p_2 ~, \\
S_{\bar{Z} \bar{Y} \bar{Z} \bar{Y}} &= -S_{\bar{Y}\bar{Z}\bar{Y}\bar{Z}} = -\tfrac{1}{2}(p_1-p_2) \frac{1+\kappa^2+p_2 (s_1-s_2+\kappa^2)}{1+\kappa^2+p_2} +\kappa^2 \frac{1+\kappa^2+p_1}{1+\kappa^2+p_2} p_2 ~,
\end{align}
\item Left-Right sector ($\varsigma_1=-1,\varsigma_2=+1$)
\begin{align}\no
S_{Z\bar{Z}Z\bar{Z}} &= -S_{Y\bar{Y}Y\bar{Y}} = \tfrac{1}{2}(p_1-p_2) \frac{1+\kappa^2+p_2 (s_1-s_2+\kappa^2)}{1+\kappa^2+p_2} +\kappa^2 \frac{1+\kappa^2-p_1}{1+\kappa^2+p_2} p_2 ~, \\
S_{Z\bar{Y}Z\bar{Y}} &= -S_{Y\bar{Z}Y\bar{Z}} = -\tfrac{1}{2}(p_1+p_2) \frac{1+\kappa^2+p_2 (s_1-s_2+\kappa^2)}{1+\kappa^2+p_2} -\kappa^2 \frac{1+\kappa^2-p_1}{1+\kappa^2+p_2} p_2 ~,
\end{align}
\item Right-Left sector ($\varsigma_1=+1,\varsigma_2=-1$)
\begin{align}\no
S_{\bar{Z}Z\bar{Z}Z} &= -S_{{\bar YY\bar YY} }-\tfrac{1}{2}(p_1-p_2) \frac{1+\kappa^2-p_2 (s_1-s_2+\kappa^2)}{1+\kappa^2-p_2} - \kappa^2\frac{1+\kappa^2+p_1}{1+\kappa^2-p_2} p_2 ~, \\
S_{{\bar{Z}Y\bar{Z}Y}} &= - S_{{\bar{Y}Z\bar{Y}Z}} = \tfrac{1}{2}(p_1+p_2) \frac{1+\kappa^2-p_2 (s_1-s_2+\kappa^2)}{1+\kappa^2-p_2} +\kappa^2 \frac{1+\kappa^2+p_1}{1+\kappa^2-p_2} p_2 ~.
\end{align}
\end{itemize}
It would be interesting to find also the fermionic sectors of this $\k$-dependent S-matrix
and check its consistency with integrability of the model, similarly to what was done for the Drinfel'd-Jimbo deformation without WZ term~\cite{Seibold:2021lju}.


\section{Concluding remarks}
\la{con}

In this paper we have constructed a family of type IIB supergravity backgrounds that are deformations of the mixed flux $\AdS_3 \times \Sp^3 \times \To^4$ background.
The ``squashed'' $\AdS_3 \times \Sp^3$ metric is naturally written in terms of Hopf fibrations and the deformed backgrounds have a number of key properties: (i) they are supersymmetric, preserving half the supersymmetry of the undeformed \adss background, (ii) they have trivial dilaton, (iii) they have regular curvature and (iv) the corresponding Green-Schwarz
superstring
sigma model is classically integrable.
Given the global symmetries and the amount of supersymmetry that is preserved, and that the fluxes are homogeneous,
one might suspect that these backgrounds may be stable under $\alpha'$-corrections 
(possibly up to redefinitions of the parameters).\foot{We should add, however, the following reservation.
While the T-duality, in general, should ``commute'' with $\a'$-corrections (modulo possible deformations
that should trivialise in half-maximally supersymmetric case)
this does not a priori apply to S-duality which is not a world-sheet symmetry.
Thus, while the original $\AdS_3 \times \Sp^3\times {\rm T}^4$ background should be stable under
$\a'$-corrections, backgrounds related to it by combinations of T- and S- dualities
may not share this property. Still, $\a'$-corrections may be absent in the special case of homogeneous fluxes.
For some discussions of the absence or presence of $\a'$-corrections to some
half-maximally supersymmetric
backgrounds with inhomogeneous fluxes see, e.g.,
\ci{Horowitz:1994ei,deHaro:2003zd}.}

The family of backgrounds \rf{8}, \rf{99}, \rf{22}--\rf{24} are deformations of $\AdS_3 \times \Sp^3$ with the $\AdS$ space written as a time-like fibration over $\Hy^2$.
The existence of this family may not be surprising since it can be found starting from the mixed flux background and T-duality and S-duality.
However, it can also be found by applying TsT transformations to the one-parameter Yang-Baxter deformation (deforming one copy of $\alg{psu}(1,1|2)$) of the mixed flux string sigma model using a particular Drinfel'd-Jimbo R-matrix.
This latter construction demonstrates the classical integrability of the corresponding string sigma model.
Let us emphasise that the DJ R-matrix used in the YB sigma model
is the unique one on $\alg{psu}(1,1|2)$ that is unimodular \cite{Hoare:2018ngg,Seibold:2019dvf}, and it
is associated
to the Dynkin diagram with all fermionic simple roots.
Crucially, this ensures that the corresponding background solves the standard type II
supergravity equations rather than the generalised supergravity equations \cite{Arutyunov:2015mqj,Wulff:2016tju}.

As discussed in the Introduction, this is another example
of S-duality unexpectedly preserving integrability, and is the first case
that involves an inhomogeneous YB deformation so that the symmetry algebra is q-deformed.
In contrast to the Jordanian examples \cite{Matsumoto:2014ubv,vanTongeren:2019dlq}, here the classical integrability does not have an alternative explanation in terms of twists and worldsheet dualities.

One motivation for studying these backgrounds is to explore the relation between integrability of bosonic sigma models and their embeddings into string theory.
In particular, given an integrable bosonic sigma model, when does there exist an embedding into supergravity (or generalised supergravity) such that the corresponding GS string sigma model is also integrable?
Moreover, if we have such a setup, when does an
S-duality transformation of type IIB background, which modifies the bosonic truncation of the worldsheet model
(leading, in general, to different metric and B-field) preserve integrability?
In all such examples that we know of, including the new ones presented here, the dilaton is constant and (setting $\Phi = \Phi_0 = 0$)
the metric is unchanged under the S-duality.

In section \ref{acb}, we considered the analytic continuation \rf{40}--\rf{39} of the family of
deformed backgrounds presented in section \ref{sec:2}
(with $\td \k= i \k$).
In this case, the $\AdS_3$ geometry is written as a space-like fibration over $\AdS_2$.
The metric interpolates between $\AdS_3 \times \Sp^3 \times \To^4$ ($\td\k = 0$) and $\AdS_2 \times \Sp^2 \times \To^6$ ($\td \k = 1$).
In the $\td \k = 1$ limit, we recover the $\AdS_2 \times \Sp^2 \times \To^6$ background with 8 supersymmetries, which for a particular choice of parameters corresponds to the near-horizon limit of the $\tfrac14$-supersymmetric intersection of four D3-branes.
It would be interesting to explore if there is any link with the 8-vertex R-matrix of \cite{deLeeuw:2021ufg} that interpolates between the building blocks of the exact $\AdS_3$ and $\AdS_2$ S-matrices, describing the scattering of string excitations above the BMN vacuum.

It is an important open question as to whether we can find a brane interpretation of the two families of deformed backgrounds.
We may expect to be in a more promising situation than the familiar $\eta$-deformation of $\AdS_5 \times \Sp^5$ \cite{Delduc:2013qra,Hoare:2018ngg} due to the special properties discussed above, in particular the trivial dilaton and supersymmetry.
The holographic interpretation of the scaling (in the analytically-continued case) background and Jordanian limit, described in sections \ref{sca} and \ref{jord} respectively, has been studied, e.g., in \cite{Alishahiha:2003ru,El-Showk:2011euy,Bena:2012wc}.
In these limits, the deformed backgrounds are naturally written in $\AdS_3$ pp-wave and Poincar\'e coordinates, and can again be generated by T-dualities and S-dualities.
It would be interesting to investigate if we can move away from these limits perturbatively, in particular in the case of the scaling background, which is
``intermediate'' between our deformed backgrounds and their Jordanian limit
(e.g., the sphere geometry is still deformed).

A brane interpretation of the analytically-continued metric in section \ref{acb}, albeit with different parameters in the $\AdS$ and $\Sp$ parts and with different supporting fluxes, has been explored in \cite{Orlando:2010ay,Orlando:2012hu}.
The backgrounds discussed there are generated by TsT transformations and the deformed brane background appears as a limit of the T-dual of the D1-D5+pp wave+KK monopole background.
The T-duality in direction transverse to KK monopole leads to ``non-geometric'' background.

A similar construction to that presented in this paper is also possible for the mixed flux $\AdS_3 \times \Sp^3 \times \Sp^3 \times \Sp^1$ background \cite{Babichenko:2009dk,Cagnazzo:2012se}.
In this case the relevant supercoset is $\frac{\grp{D}(2,1;\alpha) \times \grp{D}(2,1;\alpha)}{\grp{SU}(1,1) \times \grp{SU}(2) \times \grp{SU}(2)}$ and we can again deform a single copy of $\alg{d}(2,1;\alpha)$ preserving half the supersymmetry.
The superalgebra $\alg{d}(2,1;\alpha)$ admits a fermionic Dynkin diagram and the corresponding YB deformation should give a background solving the standard type II supergravity equations.
The $\alpha \to 0$ or $\alpha \to 1$ limits correspond to decompactifying one of the spheres, hence there should be an intersection with the deformed $\AdS_3 \times \Sp^3 \times \To^4$ backgrounds constructed here. 
Therefore, it is natural to expect that the deformed $\AdS_3 \times \Sp^3 \times \Sp^3 \times \Sp^1$ backgrounds have some of the key properties discussed above and it would be interesting to investigate the extent to which this is the case.

Finally, it would also be interesting to study string propagation on these backgrounds.
In section \rf{sec:pp} we initiated the study of the near-BMN light-cone gauge S-matrix, focusing on the limit in which the excitations become massless.
It would be interesting to move away from this limit and propose
a conjecture for the exact S-matrix along the lines of \cite{Borsato:2013qpa,Lloyd:2014bsa,Hoare:2014oua}.
It would also be instructive to study how classical solitonic string solutions, such as long strings, are affected by the deformation.

\section*{Acknowledgments}
We would like to thank M.~Cvetic, A.~Prinsloo, A.~Torrielli and L.~Wulff for useful comments.
The work of BH was supported by a UKRI Future Leaders Fellowship (grant number MR/T018909/1).
FS was supported by the Swiss National Science Foundation via the
Early Postdoc.Mobility fellowship ``q-deforming AdS/CFT''.
AAT was supported by the STFC grant ST/T000791/1.

\newpage

\appendix
\section{Details on supersymmetry}
\label{susy}

Here we shall add
the explicit form of the Killing spinor equations
in section \ref{sec:susy}
and also demonstrate that the pp-wave background of section \ref{sec:pp} preserves 16 supersymmetries.

The Killing spinor equations \rf{51} take the following explicit form
\begin{equation} \begin{aligned}\la{aa1}
& \qquad \qquad \qquad \partial_\mu \epsilon = \Omega_\mu \epsilon~,\\
4 \Omega_{\zeta_1} = \, &-(1+\kappa^2) 1_2 \otimes \Gamma^{12} - \sigma_3 \otimes (s_1 \Gamma^{12} + s_2 \Gamma^{45} ) \\
&\qquad + \sigma_1 \otimes (y_1 \Gamma^{12} + y_2 \Gamma^{45}) + i \sigma_2 \otimes (\Gamma^{12} I + \Gamma^{45} J)~, \\
4 \sqrt{1+\kappa^2} \Omega_{\bs} = \, &(1+\kappa^2) 1_2 \otimes \Gamma^{01} - \sigma_3 \otimes (s_1 \Gamma^{01} - s_2 \Gamma^{13} ) \\
&\qquad + \sigma_1 \otimes (y_1 \Gamma^{01} - y_2 \Gamma^{13}) + i \sigma_2 \otimes (\Gamma^{01} I - \Gamma^{13} J)~, \\
4 \Omega_{\xi_1} = \, & -(1+\kappa^2) 1_2 \otimes \Gamma^{45} - \sigma_3 \otimes (s_1 \Gamma^{45} + s_2 \Gamma^{12} ) \\
&\qquad + \sigma_1 \otimes (y_1 \Gamma^{45} + y_2 \Gamma^{12}) + i \sigma_2 \otimes (\Gamma^{45} I + \Gamma^{12} J)~, \\
4 \sqrt{1+\kappa^2} \Omega_{\theta} = \, & (1+\kappa^2) 1_2 \otimes \Gamma^{34} - \sigma_3 \otimes (s_1 \Gamma^{34} + s_2 \Gamma^{04} ) \\
&\qquad + \sigma_1 \otimes (y_1 \Gamma^{34} + y_2 \Gamma^{04}) + i \sigma_2 \otimes (\Gamma^{34} I + \Gamma^{04} J)~,
\\
4\Omega_{\zeta_2} = \, & \cosh \bs \Big( -(1-\kappa^2) 1_2 \otimes \Gamma^{12} + \sigma_3 \otimes (s_1 \Gamma^{12} + s_2 \Gamma^{45} ) \\
&\qquad - \sigma_1 \otimes (y_1 \Gamma^{12} + y_2 \Gamma^{45}) - i \sigma_2 \otimes (\Gamma^{12} I + \Gamma^{45} J) \Big)\\
&+\frac{\sinh \bs}{\sqrt{1+\kappa^2}} \Big( -(1+\kappa^2) 1_2 \otimes \Gamma^{02} + \sigma_3 \otimes (s_1 \Gamma^{02} - s_2 \Gamma^{23} ) \\
&\qquad - \sigma_1 \otimes (y_1 \Gamma^{02} - y_2 \Gamma^{23}) - i \sigma_2 \otimes (\Gamma^{02} I - \Gamma^{23} J) \Big)~, \\
4\Omega_{\xi_2} = \, & \cos \theta \Big( -(1-\kappa^2) 1_2 \otimes \Gamma^{45} + \sigma_3 \otimes (s_1 \Gamma^{45} + s_2 \Gamma^{12} ) \\
&\qquad - \sigma_1 \otimes (y_1 \Gamma^{45} + y_2 \Gamma^{12}) - i \sigma_2 \otimes (\Gamma^{45} I + \Gamma^{12} J) \Big)\\
&+\frac{\sin \theta}{\sqrt{1+\kappa^2}} \Big( -(1+\kappa^2) 1_2 \otimes \Gamma^{35} + \sigma_3 \otimes (s_1 \Gamma^{35} + s_2 \Gamma^{05} ) \\
&\qquad - \sigma_1 \otimes (y_1 \Gamma^{35} + y_2 \Gamma^{05}) - i \sigma_2 \otimes (\Gamma^{35} I + \Gamma^{05} J) \Big)~,
\end{aligned}
\end{equation}
where
\begin{equation}
I = y_3 \Gamma^{67} + y_5 \Gamma^{68} + y_7 \Gamma^{69}~, \qquad J = y_4 \Gamma^{67} + y_6 \Gamma^{68} + y_8 \Gamma^{69}~,
\qquad \Gamma^{a_1 \cdots a_n} \equiv \Gamma^{a_1} \cdots \Gamma^{a_n}\ .\la{aa2}
\end{equation}

In the special limit corresponding to the pp-wave background discussed in section \ref{sec:pp}
the only non-zero components of the spin connection (with curved-space indices) for the metric \rf{mm}
are
\begin{equation}
\omega_+^{-j} = - \omega_+^{j-} =\ha z_j \hat{\mu}^2~, \qquad \hat \mu =(1 + \k^2) \mu \ .
\end{equation}
For the fluxes in \rf{531}
the RR bispinor in \rf{52} is given by
\begin{equation} \begin{aligned}
\mathcal S &=- \tfrac{1}{8} \Big( -2 \mu \Gamma^{+} (\Gamma^{12}+\Gamma^{34}) \Big) \left( \sigma_1(y_1+y_2) + i \sigma_2 \big(I+J \big) \mathcal P \right)~,
\end{aligned}
\end{equation}
where $\mathcal P =\ha ( 1+\Gamma^{6789})$ is the projector involving 4-torus directions.
The Killing spinors should not depend on the torus directions. From the identities $\mathcal P \Gamma^j \mathcal P=0$ for $j=6,7,8,9$ it follows that this is automatically satisfied if $\epsilon=\mathcal P \hat{\Gamma} \epsilon$. This reduces the number of Killing spinors from $64$ to $16$.

Using that $\Gamma^+ = -\tfrac{1}{2} \Gamma_-$ and $(\Gamma^+)^2 =0$ the remaining Killing spinor equations are
\begin{align}
& \left(\partial_+ + \tfrac{1}{4}z_j \hat{\mu}^2 \Gamma_{-j}\right) \epsilon - \tfrac{1}{4} \mu(s_1+s_2) (\Gamma^{12}+\Gamma^{34}) \sigma_3 \epsilon + \mathcal S \Gamma_+ \epsilon=0~, \la{a6}\\
& \partial_- \epsilon=0 ~, \la{a7} \\
& \partial_j \epsilon - \tfrac{1}{4}\mu (s_1+s_2) \Gamma^{12+} \Gamma_j \sigma_3 \epsilon + \mathcal S \Gamma_j \epsilon=0~, \qquad j=1,2~, \la{a8}\\
& \partial_j \epsilon - \tfrac{1}{4}\mu (s_1+s_2) \Gamma^{34+} \Gamma_j \sigma_3 \epsilon + \mathcal S \Gamma_j \epsilon=0~, \qquad j=3,4~.\la{a9}
\end{align}
Eq. \rf{a7} shows that $\epsilon$ does not depend on $x^-$, while \rf{a8}, \rf{a9} imply $\partial_j \partial_k \epsilon =0$ and hence $\epsilon$ is at most linear in $z^j$. In fact, \rf{a8}, \rf{a9}
can be written as $\partial_j \epsilon = \Omega_j \epsilon$ with $\Omega_j \Omega_k =0$. The solution reads
\begin{equation}
\epsilon = (1+z^j \Omega_j) \chi~,
\end{equation}
with spinor $\chi = (\chi_1,\chi_2)$ only dependent on $x^+$. Eq. \rf{a6} then gives
\begin{equation} \begin{aligned}
(\partial_+ + \Omega_+) \chi &= -z^j \left(\com{\Omega_+}{\Omega_j} + \tfrac{1}{4}\hat{\mu}^2 \Gamma_{-j}\right) \chi = -\tfrac{1}{4}z^j \left(\hat{\mu}^2-\Vert \textbf{z}_1 + \textbf{z}_2 \Vert^2 \mu^2 \right) \Gamma_{-j} \chi ~.
\end{aligned}
\end{equation}
Both sides here must vanish. One solution is that $\Gamma_- \chi=0$, which gives 8 Killing spinors.
The other is found when
\begin{equation}
\Vert \textbf{z}_1 + \textbf{z}_2 \Vert^2 \mu^2 = \hat{\mu}^2~,
\end{equation}
which is the same as \rf{532} that was found from
the supergravity conditions \eqref{24}. This gives another set of $8$ Killing spinors.
The pp-wave background is therefore maximally supersymmetric, admitting 16 Killing spinors.

\section{Integrable YB deformation with non-unimodular Drinfel'd-Jimbo R-matrix}
\la{app:dist}
In this Appendix we present the
background corresponding to the Yang-Baxter deformed sigma model \rf{4.1} with the non-unimodular DJ R-matrix
built from a distinguished Cartan-Weyl basis.\footnote{The distinguished Dynkin diagram of $\alg{psu}(1,1|2)$ is the one with two bosonic simple roots and one fermionic simple root.}
Interpreting \rf{4.1} as the GS superstring sigma model one finds that the NSNS and RR
fluxes supporting the metric \eqref{4} are given by
\begin{equation}
\la{eq:backdist}
\begin{aligned}
H_3 &= \sqrt{1+\kappa^2} \sqrt{1-q^2} \extder \hat{B}~, \\
\mathcal F_1 &= \frac{2 \kappa q(q^2-\kappa^2)}{(q^2+\kappa^2)^{3/2}} \hat{\mathcal F}_1~, \qquad \hat{\mathcal F}_1 \equiv e^0 + e^3~, \\
\mathcal F_3 &= - \frac{\sqrt{1+\kappa^2}(q^2-\kappa^2)}{(q^2+\kappa^2)^{3/2}} \hat{\mathcal F}_3 + \frac{4 \kappa^2 q^2}{(q^2+\kappa^2)^{3/2}} \hat{\mathcal F}_1 \wedge J_2^{(1)}~, \qquad \hat{\mathcal F}_3 \equiv q^2 \extder \hat{B} - \kappa^2 \extder \check{B}~, \\
\mathcal F_5 &= - \frac{2 \kappa q \sqrt{1+\kappa^2}}{(\kappa^2+q^2)^{3/2}} \hat{\mathcal F}_3 \wedge J_2^{(1)}- \frac{ \kappa q(q^2-\kappa^2)}{(q^2+\kappa^2)^{3/2}} (1+\star) \hat{\mathcal F}_1 \wedge J_2^{(1)} \wedge J_2^{(1)}~.
\end{aligned}
\end{equation}
Here $\kappa$ and $q$ are defined as in \rf{466}, \rf{eq:yspecial} and
$\mathcal F_n $ is the analog of the combination $ e^{\Phi} F_n$ that appears in the GS action
in the standard supergravity background. The forms $e^a$, $\hat{B} , \ \check{B}$ and $J_2^{(1)}$ were defined in \rf{211}, \rf{2}, \rf{6} and \rf{14} respectively.
Note that in contrast to the supergravity background \eqref{eq:yspecial}, here the RR 1-form is non-vanishing.

When $q=1$ one recovers the pure RR deformation of \cite{Seibold:2019dvf}.\footnote{Recall that we rescale the string tension and the torus coordinates to match the deformed metric \eqref{4}.} Setting $\kappa=0$ on the other hand gives the mixed flux background \eqref{3}.

As expected from the non-unimodularity of the R-matrix, the background \rf{eq:backdist} does not solve the supergravity field equations for generic deformation parameters $\kappa$ and $q$. Instead, it satisfied generalised supergravity equations, which can be viewed as a consequence of the $\varkappa$-symmetry~\cite{Wulff:2016tju} or scale invariance of the GS sigma model~\cite{Arutyunov:2015mqj}.
There is no notion of dilaton scalar and hence it is not possible to extract standard RR field strengths $F_n$ from $\mathcal F_n$.

The generalised supergravity equations of motion are satisfied for the following choices of Killing and ``generalised dilaton'' 1-forms
(see \cite{Arutyunov:2015mqj})
\begin{equation}
I = \frac{2 \kappa q \sqrt{1+\kappa^2}}{(q^2+\kappa^2)^{3/2}}(e^0+e^3)~, \qquad Z = -\sqrt{\frac{1-q^2}{1+\kappa^2}} I~.
\end{equation}
Notice that when $\kappa=0$ or $q=0$ we have $I=Z=0$ and the background solves the standard supergravity equations of motion, with constant dilaton. $\kappa=0$ corresponds to the mixed flux background \eqref{3},
while for $q=0$ the background \eqref{eq:backdist} simplifies to (we set $\Phi=\Phi_0=0$)
\begin{equation}
\la{eq:q0}
H_3 = \sqrt{1+\kappa^2} \extder \hat{B}~, \qquad F_3 = -\kappa \sqrt{1+\kappa^2} \extder \check{B}~,
\end{equation}
which is the same as the $q=0$ case
of the supergravity DJ background \eqref{eq:yspecial}.

This can be expected from analysing the action \eqref{4.1} and, in particular, the expression \eqref{4.2}. Recalling that $q=\sqrt{1-k^2(1+\kappa^2)}$, it turns out that when $q=0$, only the term proportional to $R_g^2$ in $\Omega_-$ survives. But all DJ R-matrices obey $R^2(H_i)=0$, $R^2(E_\alpha)= - E_\alpha$, $R^2(F_\alpha)= - F_\alpha$. This no longer depends on the choice of Cartan-Weyl basis (provided the Cartan generators $H_i$ are the same). Therefore, all YB deformations for DJ R-matrices, unimodular or otherwise, will give rise to the same supergravity background \eqref{eq:q0} when $q=0$.
This is a one-parameter deformation of the pure NSNS solution, and is the S-dual of the TsT-transformed background \eqref{5}.

\medskip

Let us note that the $q=0$ case also plays a special role in the bosonic truncation of the string worldsheet sigma model
containing the parameters $\kappa, q$ and string tension $T$ (cf. \rf{4.1}).
Apart from $\ka = q = 0$ case, which corresponds to the $\grp{SL}(2,\Real)\times\grp{SU}(2)$ WZW model, this model is not conformal and $q=0$ corresponds to the fixed line of the RG flow \cite{Schubring:2020uzq}. 
The remaining parameters $\ka$ and overall scale $T$ run, such that the WZ level ${\rm k} = \sqrt{1+\ka^2}T$ is an RG invariant.\foot{Here the string tension $T$ has been rescaled compared to the action \rf{4.1}, $T \to (1+\ka^2)T$ -- see footnote \ref{trescale}.}
Moreover, this line separates the two regions with different behaviours in the UV.
A priori it is not clear why $q=0$ should correspond to a fixed line.
However, one possible explanation comes if we recall that the RG invariants of the bosonic model are \cite{Demulder:2017zhz,Schubring:2020uzq,Levine:2021fof,Klimcik:2019kkf}
\begin{equation}
{\rm k} = \sqrt{1+\ka^2}\sqrt{1-q^2} T ~, \qquad \varrho = \frac{\ka q\sqrt{1+\ka^2}\sqrt{1-q^2}}{\ka^2+q^2} ~.
\end{equation}
The deformation parameter associated to the quantum group symmetry should be an RG invariant and for $q=1$ is expected to behave as $\log {\rm q} \propto \frac{\varrho}{\rm k} = \frac{\ka}{(1+\ka^2)T}$, at least to leading order in the inverse string tension $T^{-1}$ \cite{Klimcik:2008eq,Delduc:2016ihq}.
A natural conjecture for general $q$ to leading order in $T^{-1}$ is then $\log {\rm q} \propto \frac{\varrho + O(\varrho^2)}{\rm k}$.
Now setting $q=0$ we find that $\log {\rm q} = 0$, indicating that the symmetry is not deformed and potentially explaining why $q=0$ corresponds to a fixed line of the RG flow.

A final curious observation is that for $q=0$ the proportionality constant relating the volume form of the $\AdS_3$ or $\Sp^3$ part of the metric \rf{4} and the $\AdS_3$ or $\Sp^3$ part of the 3-form flux $H$~\eqref{eq:q0} is the same as at the WZW point $\ka = q = 0$.

\section{Examples of integrable deformations with non-constant dilaton}
\la{else}

Further potentially interesting cases of integrable deformations can be constructed using other
TsT transformations in different Cartan directions.
As an example, let us start from the $q=0$ case of the DJ background \rf{eq:yspecial} (with the metric and fluxes given in \rf{4} and \rf{22} with $\hat{B}$ and $\check{B}$ defined in \rf{2} and \rf{6})
\unskip\foot{This also corresponds to the S-dual of \rf{5} or the solution \rf{244} to \rf{24} with ${\bf z}_1=( \sqrt{1+\kappa^2}, 0,0,0,0)$ and ${\bf z}_2=( 0, - \kappa \sqrt{1+\kappa^2}, 0,0,0).$}
\begin{align}
\extder s^2&= -(1+\rho^2) \extder t^2 + \frac{\extder \rho^2}{1+\rho^2} + \rho^2 \extder \psi^2 - \kappa^2 \big((1+\rho^2) \extder t - \rho^2\extder \psi\big)^2 \nn
&\qquad + (1-r^2) \extder \varphi^2 + \frac{\extder r^2}{1-r^2} + r^2 \extder \phi^2 + \kappa^2 \big((1-r^2) \extder \varphi + r^2\extder \phi\big)^2 +\extder x_s \extder x_s ~,
\label{4b} \\
\la{5b}
H_3 &= \sqrt{1+\kappa^2} \extder \hat{B} ~, \qquad \qquad F_3 = - \ka\sqrt{1+\kappa^2} \, e^{-\Phi_0} \extder \CC ~.
\end{align}
Here $\kappa =0$ corresponds to the pure NSNS background for which the bosonic part of the string action is
described by the $\grp{SL}(2,\Real)\times\grp{SU}(2)$ WZW model.
As is well known,
marginal deformations of this model can be generated using TsT
transformations in the two abelian isometry directions of $\Sp^3$ (or $\AdS_3$), corresponding to vectorial or axial gauging
in the WZW model (see, e.g., \ci{Hassan:1992gi,Giveon:1993ph}).
We can perform similar transformations also
in the case of the background \rf{4b}, \rf{5b} with $\k\neq0$.

Let us consider a TsT transformation along the two sphere isometries:
T-duality $\phi\to \td \phi $, shift $\varphi \rightarrow \varphi + \sqrt{1+\ka^2}\g \td \phi$ and T-duality back $\td \phi\to \phi $.
Similarly, in the AdS sector we may
first T-dualise $\psi\to \td \psi$, then shift $t\rightarrow t - \sqrt{1+\ka^2} \g \td \psi$ and finally T-dualise back $\td \psi\to \psi$.
To simplify the resulting geometries it will also be convenient to rescale the
isometric directions as $(\varphi,\phi,t,\psi) \rightarrow 2 (\varphi,\phi,t,\psi) $.
We then find the following metric
\begin{align} \no
\extder s^2 &= \frac{\extder \rho^2}{1+\rho^2} +\frac{1}{h_\rho} \big( -(1+\rho^2) \extder t^2 + \rho^2 \extder \psi^2 - \kappa^2 \big((1+\rho^2) \extder t - \rho^2 \extder \psi\big)^2 \big) \\
& \qquad + \frac{\extder r^2}{1-r^2} + \frac{1}{h_r} \big( (1-r^2) \extder \varphi^2 + r^2 \extder \phi^2 + \kappa^2 \big((1-r^2) \extder \varphi + r^2 \extder \phi\big)^2 \big) +\extder x_s\extder x_s\,,
\la{b44}\\
h _\rho &= \tfrac{1}{16} \big((\g+2)^2(1+\rho^2) -(\g-2)^2\rho^2\big)\,,\qquad
\quad h _r = \tfrac{1}{16} \big( (\g+2)^2(1-r^2) + (\g-2)^2 r^2\big)\,.
\no
\end{align}
The $H_3$ flux and the dilaton become
\begin{align} \la{b55}
H_3 &= \tfrac14 \sqrt{1+\ka^2} (\g^2 -4) \big( \frac{\rho^2}{h_\rho^2}
\extder \rho\wedge \extder t \wedge \extder \psi
+ \frac{ r^2}{h_r^2} \extder r \wedge \extder \varphi \wedge \extder \phi \big) \ , \\
\Phi &= \Phi_0 -\tfrac{1}{2} \log \big( h_r h_\rho\big)\,,
\end{align}
while the $F_3$ flux remains unchanged, i.e.~is the same as in \rf{5b}, up to the rescaling of the isometric directions.

Note that the $H_3$ flux vanishes when
$\g=\pm 2$.
In the undeformed theory ($\kappa=0$) for the $\Sp^3$ part these two special values of
$\g$ correspond to the axially-gauged and the vectorially-gauged $\frac{\grp{SU}(2)}{\grp{U}(1)} \times \grp{U}(1)$ gauged WZW models respectively.
For $\k\neq0$, setting $\g=2 $ we have $h_\rho= 1 + \rho^2$ and $h_r= 1 - r^2$ so that the metric in \rf{b44} becomes
\begin{equation} \begin{aligned}
\extder s^2 &= -\extder t^2 + \frac{\extder \rho^2}{1+\rho^2} + \frac{\rho^2}{1+\rho^2} \extder \psi^2- \frac{\kappa^2}{1+\rho^2} \big((1+\rho^2) \extder t-\rho^2 \extder \psi\big)^2 \\
&\qquad + \extder \varphi^2 + \frac{\extder r^2}{1-r^2} + \frac{r^2}{1-r^2} \extder \phi^2+ \frac{\kappa^2}{1-r^2} \big((1-r^2) \extder \varphi+ r^2 \extder \phi\big)^2 + \extder x_s \extder x_s~.
\end{aligned}
\end{equation}
Its Ricci scalar is $R=-4 (1+\kappa^2) \frac{r^2+\rho^2}{(1-r^2)(1+\rho^2)}$. The dilaton is
$\Phi = \Phi_0-\frac{1}{2} \log \big((1+\rho^2)(1-r^2) \big)$ and
we also have the non-zero RR $F_3$ flux in \rf{5b}.
For $\g=-2$ we find the metric
\begin{equation} \begin{aligned}
\extder s^2 &= \frac{1+\rho^2}{\rho^2}\extder t^2 + \frac{\extder \rho^2}{1+\rho^2} -\extder \psi^2+\frac{\kappa^2}{\rho^2} \big((1+\rho^2) \extder t-\rho^2 \extder \psi\big)^2\\
&\qquad + \frac{1-r^2}{r^2}\extder \varphi^2 + \frac{\extder r^2}{1-r^2} + \extder \phi^2+ \frac{\kappa^2}{r^2} \big((1-r^2) \extder \varphi+ r^2 \extder \phi\big)^2 + \extder x_s \extder x_s~,
\end{aligned}
\end{equation}
with the Ricci scalar $R = -4 (1+\kappa^2) \frac{r^2+\rho^2}{r^2 \rho^2}$. The dilaton is
$
\Phi = \Phi_0-\frac{1}{2} \log \left(-\rho^2 r^2 \right) $
implying that $\Phi_0$ should be shifted by an imaginary constant, which makes $F_3$ in \rf{5b} imaginary.
For $\g \geq 0$ the function $h_\rho$ is strictly positive, while for $\g < 0$ it is negative if $\rho^2 > -\frac{(2+\g)^2}{8\g}$.
This suggests either restricting $\g$ to be positive or
that the solution needs to be analytically continued for negative $\g$.

\bibliographystyle{nb}
\bibliography{HST}

\end{document}

\begin{bibtex}[\jobname]

\end{bibtex}

\bibliographystyle{nb}
\bibliography{\jobname}

\end{document}